\documentclass[twocolumn]{aastex7}

\newcommand{\MBH}{M_{\scalebox{1.2}{$\bullet$}}}
\newcommand{\Mstar}{M_{\bigstar}}
\usepackage{hyperref}
\usepackage{comment}
\usepackage{xcolor}

\begin{document}



\title{Beyond the Dot: an LRD-like nucleus at the Heart of an IR-Bright Galaxy and its implications for high-redshift LRDs}

\author[orcid=0000-0002-5104-8245,sname='Pierluigi Rinaldi']{Pierluigi Rinaldi}
\affiliation{Steward Observatory, University of Arizona, 933 North Cherry Avenue, Tucson, AZ 85721, USA}
\email[show]{prinaldi@arizona.edu}  

\author[orcid=0000-0003-2303-6519,sname='George H. Rieke']{George H. Rieke}
\email[]{}  
\affiliation{Steward Observatory, University of Arizona, 933 North Cherry Avenue, Tucson, AZ 85721, USA}

\author[orcid=0000-0002-8876-5248,sname='Zihao Wu']{Zihao Wu}
\email[]{}  
\affiliation{Center for Astrophysics $\vert$ Harvard \& Smithsonian, 60 Garden St., Cambridge MA 02138 USA}

\author[orcid=0009-0001-4012-3043,sname='Carys J. E. Gilbert']{Carys J. E. Gilbert}
\email[]{}  
\affiliation{Department of Astronomy, University of Cape Town, Private Bag X3, Rondebosch 7701, South Africa}

\author[orcid=0000-0001-9879-7780,sname='Fabio Pacucci']{Fabio Pacucci}
\email[]{}  
\affiliation{Center for Astrophysics $\vert$ Harvard \& Smithsonian, 60 Garden St., Cambridge MA 02138 USA}
\affiliation{Black Hole Initiative, Harvard University, 20 Garden St, Cambridge, MA 02138, USA}

\author[orcid=0000-0003-3419-538X,sname='Luigi Barchiesi']{Luigi Barchiesi}
\email[]{}  
\affiliation{Department of Astronomy, University of Cape Town, Private Bag X3, Rondebosch 7701, South Africa}
\affiliation{Inter-University Institute for Data Intensive Astronomy, University of Cape Town, Private Bag X3, Rondebosch 7701, South Africa}
\affiliation{INAF–Istituto di Radioastronomia, Via Piero Gobetti 101, I-40129 Bologna, Italy}

\author[orcid=0000-0002-8909-8782,sname='Stacey Alberts']{Stacey Alberts}
\email[]{} 
\affiliation{AURA for the European Space Agency (ESA), Space Telescope Science Institute, 3700 San Martin Dr., Baltimore, MD 21218, USA}
\affiliation{Steward Observatory, University of Arizona, 933 North Cherry Avenue, Tucson, AZ 85721, USA}

\author[orcid=0000-0002-6719-380X,sname='Stefano Carniani']{Stefano Carniani}
\email[]{} 
\affiliation{Scuola Normale Superiore, Piazza dei Cavalieri 7, I-56126 Pisa, Italy}

\author[orcid=0000-0002-8651-9879,sname='Andrew J. Bunker]{Andrew J. Bunker}
\email[]{} 
\affiliation{Department of Physics, University of Oxford, Denys Wilkinson Building, Keble Road, Oxford OX1 3RH, UK}

\author[orcid=0000-0003-0883-2226,sname='Rachana Bhatawdekar']{Rachana Bhatawdekar}
\email[]{}
\affiliation{European Space Agency (ESA), European Space Astronomy Centre (ESAC), Camino Bajo del Castillo s/n, 28692 Villanueva de la Cañada, Madrid, Spain}

\author[orcid=0000-0003-2388-8172,sname='Francesco D'Eugenio']{Francesco D'Eugenio}
\email[]{} 
\affiliation{Kavli Institute for Cosmology, University of Cambridge, Madingley Road, Cambridge, CB3 0HA, UK}
\affiliation{Cavendish Laboratory, University of Cambridge, 19 JJ Thomson Avenue, Cambridge, CB3 0HE, UK}

\author[orcid=0000-0001-7673-2257,sname='Zhiyuan Ji']{Zhiyuan Ji}
\email[]{} 
\affiliation{Steward Observatory, University of Arizona, 933 North Cherry Avenue, Tucson, AZ 85721, USA}

\author[orcid=0000-0002-9280-7594,sname='Benjamin D.\ Johnson']{Benjamin D.\ Johnson}
\email[]{} 
\affiliation{Center for Astrophysics $|$ Harvard \& Smithsonian, 60 Garden St., Cambridge MA 02138 USA}

\author[orcid=0000-0003-4565-8239,sname='Kevin Hainline']{Kevin Hainline}
\email[]{} 
\affiliation{Steward Observatory, University of Arizona, 933 North Cherry Avenue, Tucson, AZ 85721, USA}

\author[orcid=0000-0002-5588-9156,sname='Vasily Kokorev']{Vasily Kokorev}
\email[]{}
\affiliation{Department of Astronomy, The University of Texas at Austin, Austin, TX 78712, USA}

\author[orcid=0000-0002-5320-2568,sname='Nimisha Kumari']{Nimisha Kumari}
\email[]{}
\affiliation{AURA for the European Space Agency (ESA), Space Telescope Science Institute, 3700 San Martin Dr., Baltimore, MD 21218, USA}

\author[orcid=0000-0001-8386-3546,sname='Edoardo Iani']{Edoardo Iani}
\email[]{}

\affiliation{Institute of Science and Technology Austria (ISTA), Am Campus 1, 3400 Klosterneuburg, Austria}

\author[orcid=0000-0002-6221-1829,sname='Jianwei Lyu']{Jianwei Lyu}
\email[]{}  
\affiliation{Steward Observatory, University of Arizona, 933 North Cherry Avenue, Tucson, AZ 85721, USA}

\author[orcid=0000-0002-4985-3819,sname='Roberto Maiolino']{Roberto Maiolino}
\email[]{} 
\affiliation{Kavli Institute for Cosmology, University of Cambridge, Madingley Road, Cambridge, CB3 0HA, UK}
\affiliation{Cavendish Laboratory, University of Cambridge, 19 JJ Thomson Avenue, Cambridge, CB3 0HE, UK}
\affiliation{Department of Physics and Astronomy, University College London, Gower Street, London WC1E 6BT, UK}

\author[orcid=0000-0002-7392-7814,sname='Eleonora Parlanti']{Eleonora Parlanti}
\email[]{} 
\affiliation{Scuola Normale Superiore, Piazza dei Cavalieri 7, I-56126 Pisa, Italy}

\author[orcid=0000-0002-4271-0364,sname='Brant Robertson']{Brant E. Robertson}
\email[]{}
\affiliation{Department of Astronomy and Astrophysics, University of California, Santa Cruz, 1156 High Street, Santa Cruz, CA 95064, USA}

\author[orcid=0000-0001-6561-9443,sname='Yang Sun']{Yang Sun}
\email[]{}
\affiliation{Steward Observatory, University of Arizona, 933 North Cherry Avenue, Tucson, AZ 85721, USA}

\author[orcid=0000-0002-8853-9611,sname='Cristian Vignali']{Cristian Vignali}
\email[]{} 
\affiliation{Dipartimento di Fisica e Astronomia, Alma Mater Studiorum, Università degli Studi di Bologna, Via Gobetti 93/2, 40129 Bologna,
Italy}
\affiliation{INAF–Osservatorio di Astrofisica e Scienza dello Spazio di Bologna, Via Gobetti 93/3, 40129 Bologna, Italy}

\author[orcid=0000-0003-2919-7495,sname='Christina C. Williams']{Christina C. Williams}
\email[]{} 
\affiliation{NSF National Optical-Infrared Astronomy Research Laboratory, 950 North Cherry Avenue, Tucson, AZ 85719, USA}

\author[orcid=0000-0001-9262-9997,sname='Christopher  N.\ A.\ Willmer']{Christopher  N.\ A.\ Willmer}
\email[]{} 
\affiliation{Steward Observatory, University of Arizona, 933 North Cherry Avenue, Tucson, AZ 85721, USA}

\author[orcid=0000-0003-3307-7525,sname='Yongda Zhu']{Yongda Zhu}
\email[]{} 
\affiliation{Steward Observatory, University of Arizona, 933 North Cherry Avenue, Tucson, AZ 85721, USA}

\begin{abstract}

Little Red Dots (LRDs) are compact, red sources discovered by JWST at high redshift ($z \gtrsim 4$), marked by distinctive “V-shaped” spectral energy distributions (SEDs) and often interpreted as rapidly accreting Active Galactic Nuclei (AGNs).  Their true nature remains unclear though, and their evolutionary connection to their lower-redshift counterparts is still poorly constrained. Thus, we present WISEA J123635.56+621424.2 (here dubbed {\it the Saguaro}), a $z=2.0145$ galaxy in GOODS-North, as a possible analog of high-redshift LRDs and a potential missing link in their evolutionary path toward lower-redshift systems. It features a compact LRD-like nucleus surrounded by a face-on spiral host.  Its connection to LRDs includes that: (1) its nuclear spectrum shows a clear “V-shaped” SED; and (2) when redshifted to $z=7$, surface brightness dimming makes the host undetectable, thus mimicking an LRD. This suggests that high-redshift LRDs may be embedded in extended hosts. To test this, we stack rest-frame UV images of 99 photometrically selected LRDs, revealing faint, diffuse emission. Stacking in redshift bins reveals mild radial growth, consistent with the expected galaxy size evolution. A simple analytic model confirms that surface brightness dimming alone can explain their compact appearance. Lastly, we show that {\it the Saguaro} is not unique by describing similar objects from the literature at $z\lesssim3.5$. Taken together, our results support a scenario in which LRDs may not be a distinct population, but could be the visible nuclei of galaxies undergoing a short-lived, (perhaps)  AGN-dominated evolutionary phase, with their compact, red appearance driven largely by observational biases.

\end{abstract}

\keywords{Active galactic nuclei (16); High-redshift galaxies (734); Galaxy evolution (594); Near in-frared astronomy (1093); AGN host galaxies (2017); Galaxy formation (595); Photoionization (2060); Spectral energy distribution (2129);  Infrared astronomy (786); Galaxies (573); Infrared photometry (792)}


\section{Introduction}

One of the goals in building JWST (\citealt{gardner_james_2023}) was to see far enough back in time that fundamental differences would be apparent in galaxies and active galactic nuclei (AGNs). Pre-JWST, although number counts of nearly all classes of objects were found to evolve strongly, the properties of individual objects seemed familiar in comparison with relatively nearby analogs (e.g., for quasars \citealt{fan_quasars_2023}, for lower luminosities \citealt{padovani_active_2017}). This implied that we were not yet reaching far enough back to witness their birth and earliest stages of evolution.

JWST has now extended our view well into the infrared, providing the sensitivity and wavelength coverage needed to study AGN populations deep into the early Universe.

For example, Seyfert-luminosity AGNs at $z > 4$ appear to host significantly overmassive supermassive black holes (SMBHs; e.g., \citealt{harikane_jwstnirspec_2023, maiolino_jades_2023, pacucci_jwst_2023}). Even more interestingly, JWST has  unveiled an apparently new class of compact, red sources, commonly referred to as “{\it Little Red Dots}” (LRDs; e.g., \citealt{furtak_jwst_2023, killi_deciphering_2023, kokorev_uncover_2023, ubler_ga-nifs_2023, akins_cosmos-web_2024, barro_extremely_2024, greene_uncover_2024, kocevski_rise_2024, kokorev_census_2024, 
matthee_little_2024,
hainline_investigation_2024, perez-gonzalez_what_2024, rinaldi_not_2025, williams_galaxies_2024, rinaldi_way_2026}).

These sources have attracted significant attention because of their puzzling properties. For instance, they are compact and red at rest-frame optical wavelengths, yet blue in the UV, where they often display complex morphologies that may indicate merger activity (\citealt{rinaldi_not_2025}). They show the characteristic “V-shaped” spectral energy distribution (SED), with a turnover typically occurring around the Balmer limit (\citealt{setton_little_2024}). These sources defy classical templates of star-forming galaxies and AGNs (see discussion in, e.g.,  \citealt{de_graaff_remarkable_2025, ji_blackthunder_2025}). They also show broad permitted emission lines, typically interpreted as evidence for AGN activity (e.g., \citealt{maiolino_jades_2023, pacucci_jwst_2023}). Under this interpretation, many studies suggest that LRDs host overmassive black holes relative to their galaxies ($\MBH \approx 10^{6-8}\;M_{\odot}$; e.g., \citealt{furtak_jwst_2023, killi_deciphering_2023, kokorev_uncover_2023, pacucci_jwst_2023, matthee_little_2024, rinaldi_not_2025, zhang_abundant_2025}), challenging models of early black hole growth.  However, their role and relative importance in AGN evolution remain uncertain, especially because their luminosities and black hole mass estimates are still highly uncertain. 

An alternative interpretation is the recently proposed “Black Hole Star” (BH*) scenario (e.g., \citealt{naidu_black_2025, kido_black_2025, liu_balmer_2025}), in which LRDs are powered by an unknown central engine embedded in a dense, partly ionized, high-column-density gas layer ($n_{\rm H}\approx10^{8\text{--}10},\mathrm{cm}^{-3}$; $N_{\rm H}\approx10^{23\text{--}24},\mathrm{cm}^{-2}$). This configuration may explain the strong Balmer breaks, Balmer absorption, and intense Balmer emission observed in some LRDs (e.g., \citealt{naidu_black_2025, ji_blackthunder_2025, de_graaff_little_2025, de_graaff_remarkable_2025}). In this framework, broad Balmer lines can arise from exponential profiles produced by electron scattering and radiative-transfer effects, reducing the inferred black-hole masses and easing the tension with local black-hole--stellar-mass relations (e.g., \citealt{rusakov_little_2026, naidu_black_2025}). However, no coherent picture has yet emerged, and the nature of LRDs remains unclear. 

Furthermore, they are typically X-ray and radio faint \citep{ananna_x-ray_2024, mazzolari_radio_2024, yue_stacking_2024, maiolino_jwst_2025, perger_deep_2025}, although a few X-ray detections have been reported \citep{kocevski_rise_2025, hviding_x-ray_2026} and some radio-detected cases also exist \citep{zhong_blackbody_2026}. Their infrared properties are equally puzzling: while some studies report warm/hot-dust signatures (e.g., \citealt{barro_cliff_2025}), others find a deficit at these wavelengths (e.g., \citealt{williams_galaxies_2024}). However, MIRI-based stacking analyses suggest that, on average, LRDs do show warm/hot-dust emission (e.g., \citealt{delvecchio_active_2025}).

Finally, LRDs appear abundant at $z > 4$ ($\approx100$ times more numerous than quasars at similar $M_{\rm UV}$ and redshift; \citealt{kokorev_census_2024, akins_cosmos-web_2025, kocevski_rise_2025, rinaldi_way_2026}), yet decline rapidly at $z < 4$ (e.g., \citealt{bisigello_euclid_euclid_2025, ma_counting_2025}). However, recent studies have suggested that selection biases may play an important role in shaping the inferred demographics of LRDs across cosmic time (e.g., \citealt{rinaldi_way_2026}). Together with the diverse properties observed in these sources, this suggests that LRDs may represent a far more heterogeneous population than initially thought. Despite the large number of studies published since their discovery, their true nature remains debated, requiring a multi-faceted approach to characterize them. Given their likely heterogeneous nature and the uncertainty surrounding their physical origin, throughout this work we refer to LRDs as photometrically selected compact, red sources with a “V-shaped” SED.

In this work, we want to address two fundamental questions: {\it what do these sources evolve into as the Universe matures?} and {\it do they emerge from ordinary galaxies, or are they born in fundamentally different environments?} Answering the first question is  difficult as observations have shown that the number density of LRDs drops dramatically toward lower redshifts (\citealt{kocevski_rise_2024, kokorev_census_2024, ma_counting_2025}). They may evolve into a variety of objects with related, but different characteristics, making identifications ambiguous. The scarcity of LRDs at $z < 4$ also makes it difficult to answer the second question, as it limits our ability to study analogs in a regime where host galaxies would be more easily detected and used to inform the nature of their high-redshift counterparts.

In this paper, we will look at relatively low redshift analogs both to learn more about the possible family tree of canonical LRDs and to see if they give hints on how to interpret the observations at higher redshift. LRDs have been identified at $z \lesssim 3-4$, extending this population to later cosmic times (e.g., \citealt{juodzbalis_jades_2024, lin_discovery_2024, stepney_big_2024, loiacono_big_2025, de_graaff_remarkable_2025, ma_counting_2025, hviding_x-ray_2026}). Nonetheless, these discoveries are rare and no clear picture has emerged. Fortunately, rapid progress is being made to find larger samples. Using a thorough search of the SDSS database, \citet{lin_discovery_2025} have identified three objects at $z \approx 0.1$ that match LRD properties closely (see also \citealt{ji_lord_2026}). This is an important advance, but emphasizes  how rare local close matches to canonical LRDs are. Furthermore, the search for low-$z$ counterparts has recently been extended to the Dark Energy Spectroscopic Instrument (DESI; \citealt{collaboration_data_2026}) survey (see \citealt{ding_discovery_2026, park_new_2026, lin_lrds2_2026}).  

{ Recently,} \citet{billand_investigating_2025} have explored the possibility that LRDs evolve by forming  envelopes of young extended galaxies. To explore this possibility, they selected from a much larger sample of 55 red galaxies at $z_{med} = 3.6 \pm 1.1$ with such envelopes and  derived their  properties, concluding that these objects likely represent the next evolutionary step. However, their study does not explore alternative pathways: an important limitation, given the broad range of properties observed in canonical LRDs at high redshifts (see discussion in \citealt{perez-gonzalez_little_2026}). { That is, both their physical nature and, most importantly, their evolutionary fate at later cosmic times remain unclear.}

{ In this paper, we discuss the missing link between canonical LRDs at high redshift and their counterparts at later cosmic times, namely objects potentially related to LRDs at $0.2 < z < 3.5$.} Many of the proposed LRD analogs have been identified through JWST observations. It is important to note that, prior to JWST, the {\it Spitzer} mission (\citealt{gehrz_nasa_2007}) identified a variety of extragalactic sources in this redshift regime, the so-called Blue-excess Dust-obscured Galaxies (Blue HotDOG; \citealt{assef_hot_2016, noboriguchi_extreme_2022}), which exhibit SEDs remarkably similar to those of LRDs as highlighted by \citet{noboriguchi_similarity_2023}. Thus, they are a second good hunting ground for LRD analogs.

As part of the ongoing effort to bridge this gap, we have studied in detail WISEA J123635.56+621424.2 (here dubbed {\it the Saguaro}\footnote{In reference to the rare spiral-shaped crested Saguaros found in the Arizona desert; both are strikingly sculptural, with some of them being clearly spiral in form, and span impressive scales in their respective environments.}), originally brought to prominence with {\it Spitzer} as an infrared bright, optically faint galaxy \citep[e.g.,][]{donley_agn_2010}.  It is a well-studied object at $z = 2.0145$, located in GOODS-North (GOODS-N). It has an obscured AGN and a compact core in imaging from the {\it Hubble} Space Telescope (HST). Furthermore, its X-ray flux is very faint, indicating that it is Compton Thick (CT) or nearly so (\citealt{donley_agn_2010}). Data from the Near-Infrared Spectrograph (NIRSpec; \citealt{ferruit_near-infrared_2022, jakobsen_near-infrared_2022}) show the classical “V-shape” feature at its core, similar to that seen in canonical LRDs. While its integrated SED differs somewhat from LRDs, its nuclear SED shows striking similarity, suggesting it may represent a more evolved descendant or an analog.

Notably, {\it the Saguaro} consists of a compact LRD-like nucleus {\it plus} a face on spiral galaxy surrounding it. This provides an opportunity to explore whether, at high redshift, it would appear as a compact LRD, or whether it would be disqualified by the surrounding structure. We explore this question both by synthetically moving it to z $\approx$ 7 and also by an analysis of the impact of surface brightness dimming on the detectability of LRD host galaxies in general.

\vspace{2mm}

This paper is organized as follows. In Section 2, we introduce {\it the Saguaro} and present its main observational characteristics. We then highlight its resemblance to canonical LRDs through three independent lines of evidence: (1) multiwavelength AGN–host image decomposition; (2) a spatially resolved NIRSpec/PRISM spectrum in its nuclear regions; and (3) a redshifting experiment illustrating how the system would appear if observed at higher redshifts. In Section 3, building on the findings from Section 2, we investigate, through stacking analysis, whether LRDs at $z \approx 4\text{--}8$ may similarly be embedded within faint extended emission that becomes increasingly suppressed with redshift due to cosmological surface brightness dimming. We then develop a simple analytic model quantifying the expected loss of detectable light from LRDs at increasing redshifts. In Section 4, we show that {\it the Saguaro} is not an isolated case. Finally, in Section 5, we discuss and summarize our findings.

\vspace{2mm}
Throughout this paper, we consider a cosmology with $H_{0} = 70\; \rm km\;s^{-1}\;Mpc^{-1}$, $\Omega_{M} = 0.3$, and $\Omega_{\Lambda} =0.7$. All magnitudes are total and refer to the AB system \citep{oke_secondary_1983}. A \citet{kroupa_variation_2001} initial mass function (IMF) is assumed (0.1--100 M$_{\odot}$).

\section{{\it the Saguaro}: A Case Study}

\begin{figure*}
    \centering
    \includegraphics[width=1.0\linewidth]{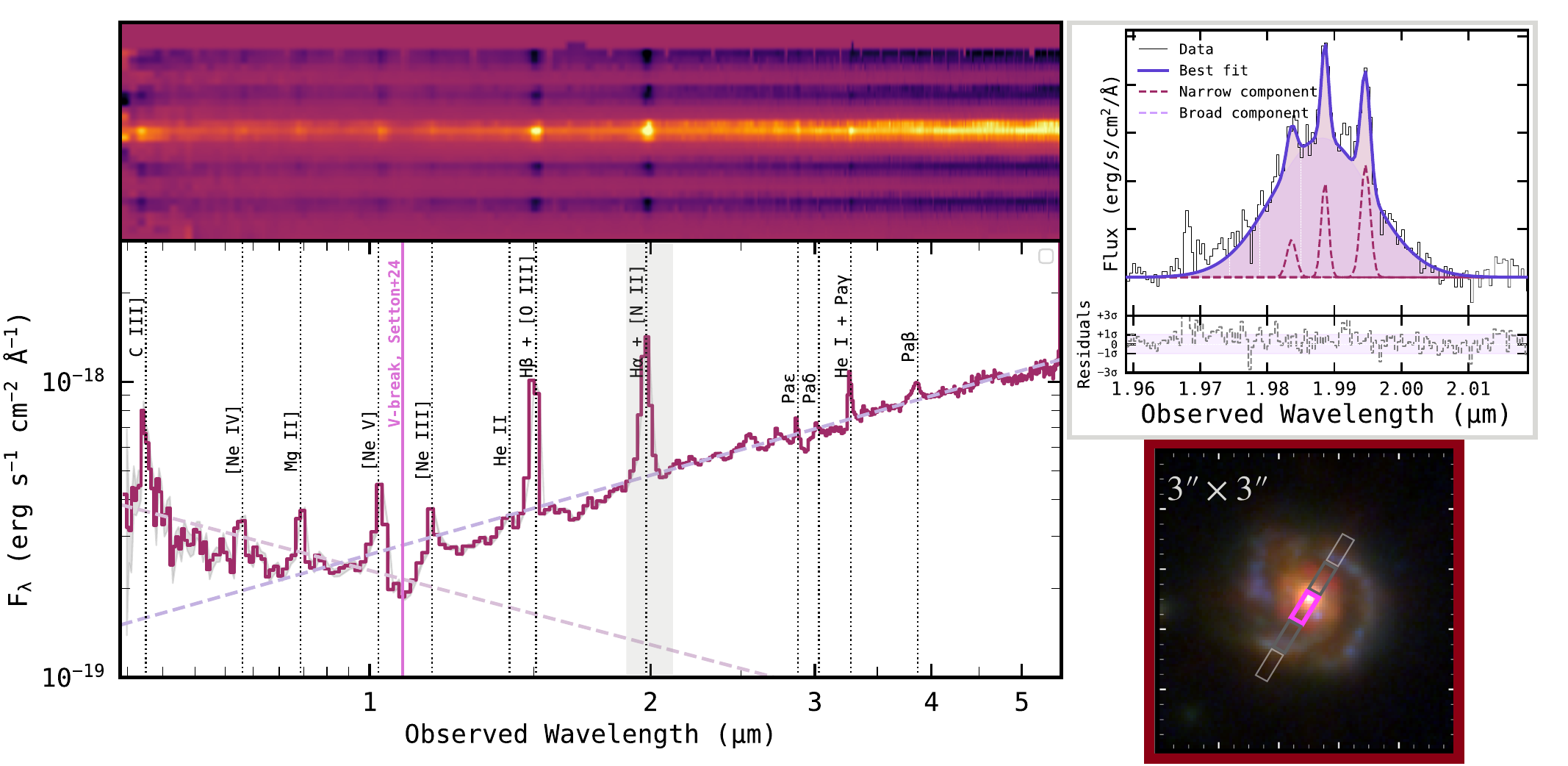}

\includegraphics[width=1.0\linewidth]{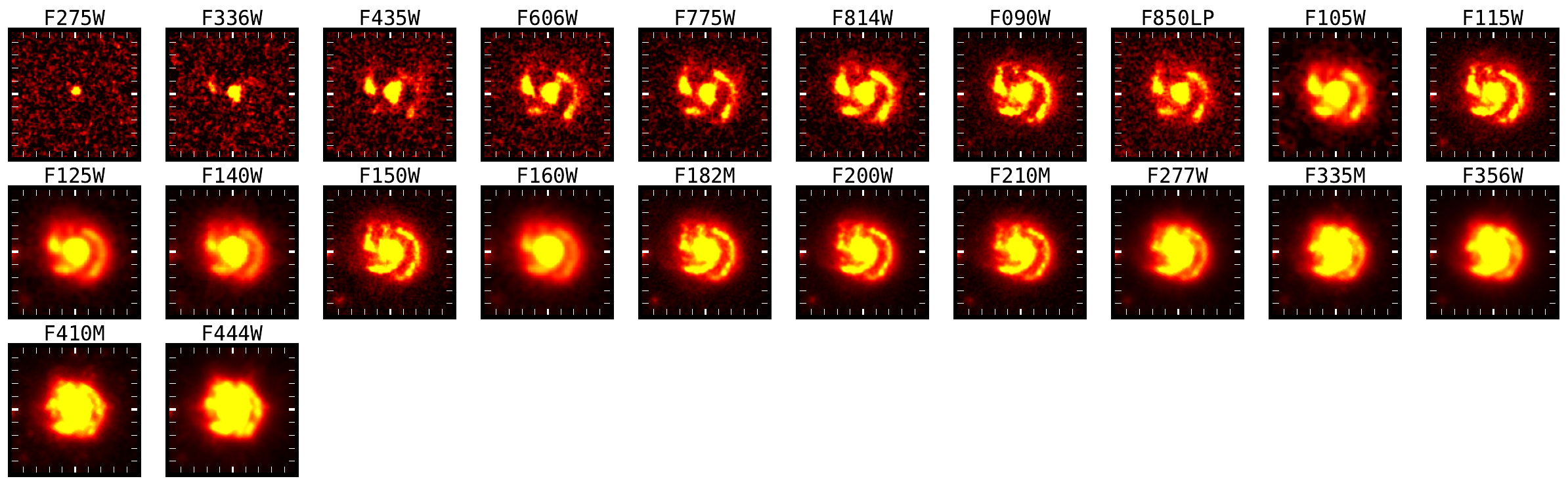}
    \caption{{ Top:} On the left, we show the NIRSpec/PRISM spectrum of {\it the Saguaro} (from the DJA DAWN Archive; v4.4), with all prominent emission lines labeled. The expected position of the { break}, as defined by \citet{setton_little_2024}, is marked in orchid. { We also note the presence of [Ne\,{\sc iv}]$\lambda\lambda2422,2424$, as already reported in higher-$z$ LRDs (see, e.g., \citealt{tang_spurs_2026}).} The H$\alpha$ + [N\,{\sc ii}]$\lambda\lambda6548,6583$ complex (unresolved in the PRISM spectrum) is highlighted in gray. Its higher-resolution counterpart from the NIRSpec G235H/F170LP grating is shown in the top-right panel, along with the line model fits for both H$\alpha$ and [N\,{\sc ii}]$\lambda\lambda6548,6583$ { (although we highlight the limited data quality of this dataset)}. Finally, we show the NIRCam RGB image (3\arcsec$\times$3\arcsec) of {\it the Saguaro} with the slit overlaid, illustrating that the PRISM flux originates from the nuclear regions. { Bottom:} Postage stamps (3\arcsec$\times$3\arcsec) of the available HST and JWST filters for {\it the Saguaro}.}
    \label{fig:wise}
\end{figure*}

The characteristic features of high-redshift LRDs are their “V-shaped” SEDs, typically emerging around { the Balmer limit} (\citealt{setton_little_2024}), and their compactness in { NIRCam/F444W}. { They often show broad Balmer lines, commonly interpreted as evidence for active black hole accretion, although this interpretation remains actively debated.} While this distinctive spectral shape has been predominantly observed at $z\gtrsim 3-4$, recent efforts have uncovered a few rare cases at lower redshifts (e.g., \citealt{wang_rubies_2024,  juodzbalis_jades_2024}). We propose another intriguing example at $z\approx2$: {\it the Saguaro}, located in the GOODS-N field ($\mathrm{R.A.} = 189.1482812$, $\mathrm{Dec} = 62.2400236$), with an extensive multi-wavelength coverage ranging from X-ray to submillimeter wavelengths.


In this section, we present {\it the Saguaro} and describe its general properties. We show that it has a “V-shaped” SED, which arises from the bright central core of the galaxy, based on a multi-wavelength AGN–host image decomposition and a spatially resolved analysis of the NIRSpec/PRISM spectrum in its nuclear region. Finally, we investigate how this object would appear if observed at high redshifts, revealing a striking resemblance to canonical LRDs.

\subsection{Dataset}
We used both imaging and spectroscopic data from the GOODS-N field, obtained with the {\it Hubble} Space Telescope (HST) and JWST. 

For HST, we relied on the {\it Hubble} Legacy Field (HLF) dataset, which provides broad wavelength coverage from 0.2 to 1.6~$\mu$m, including UV bands (WFC3/UVIS: F275W and F336W), optical bands (ACS/WFC: F435W, F606W, F775W, F814W, F850LP) and near-infrared bands (WFC3/IR: F105W, F125W, F140W, F160W). A detailed description is available in \citet{whitaker_hubble_2019}\footnote{HLF imaging is publicly available at \url{https://archive.stsci.edu/prepds/hlf/}.}.

For JWST, we used both NIRCam and NIRSpec observations. NIRCam imaging was obtained from JADES/NIRCam Data Release 2 (DR2; PIDs: 1181; PIs: D. Eisenstein; \citealt{eisenstein_overview_2023, eisenstein_jades_2023, deugenio_jades_2024}\footnote{JADES images available at \url{https://archive.stsci.edu/hlsp/jades}.}), complemented by data from the FRESCO program (PID: 1895; PI: P. Oesch; \citealt{oesch_jwst_2023}). These images reach 5$\sigma$ depths of 29.3–29.9~mag (measured in 0.15\arcsec-radius apertures; see \citealt{deugenio_jades_2024} for more details on the data reduction). NIRSpec observations were obtained as part of the NIRSpec GTO WIDE survey (PID: 1211, PI: K. Isaak; \citealt{maseda_nirspec_2024}). { We made use of both PRISM and high resolution data (G235H/F170LP and G395H/F290LP), the latter used to showcase the presence of a broad H$\alpha$ and absorption feature in He\,{\sc i}$\lambda10830$. For the PRISM observations, we primarily adopted the spectrum from the DAWN JWST Archive (DJA; see Figure~\ref{fig:wise} and Figure~\ref{fig:agn-host-dec}). However, for the spatial analysis presented in Section~\ref{sec:nuclear_v-shape}, we used our own reduction of the PRISM data with {\sc MSAEXP}\footnote{\url{https://github.com/gbrammer/msaexp}} ({\tt v0.9.8}; \citealt{brammer_msaexp_2023}). This dedicated reduction is used exclusively for that analysis, since the default nodding subtraction adopted in the DJA products can introduce self-subtraction effects that compromise a reliable spatial investigation of the nuclear region. The G235H/F170LP and G395H/F290LP data were fully reduced with {\sc MSAEXP} from the outset.}\footnote{{ We note that Saguaro is covered by both G235H/F170LP and G395H/F290LP. However, both datasets are affected by cosmic rays and detector artifacts, and the limited number of exposures prevents their reliable use. Future high-resolution observations of this source will soon become available as part of PID 10311 (PIs: P. Rinaldi \& E. Iani).}}


{\it The Saguaro} is also detected in the deep {\it Chandra} 2Ms X-ray data (\citealt{xue_2_2016}; {\tt ID}: CDFN$-190$) with $L_{X,\,2-10\,\text{keV, int}} = (6.61\text{--}10)\times10^{43}\,\rm erg\,s^{-1}$ (\citealt{donley_agn_2010, del_moro_mid-infrared_2016}), with an absorbing column density of $N_H \approx 10^{23.6-23.8}\,\text{cm}^{-2}$(\citealt{alexander_x-ray_2005, donley_agn_2010}).  
The source shows no detection in the Nuclear Spectroscopic Telescope Array (NuSTAR) data (\citealt{harrison_nuclear_2013}). To validate the previous findings in the X-ray, we re-extracted and analysed all the {\it Chandra} observations using the Chandra Interactive Analysis of Observations (CIAO; v4.17 and CALDB 4.11.0; \citealt{fruscione_ciao_2006}) tools ({\tt specextract}). The spectra from different observations (covering a span of 1.5 years) were combined in a single one via the \textsc{CIAO} tool \texttt{combine\_spectra}. We did not find any significant variability in the covered time range.
The object has an effective coverage of $\approx1.96\,\rm{Ms}$, with 246 net-counts (i.e. background subtracted). Given the number of counts, we performed the spectral analysis using unbinned data and \textit{C}-statistic \citep{cash_parameter_1979}. {Due to the high background counts, we restricted our spectral fitting to rest-frame energies $E>3.5\,\rm{keV}$, i.e., where the spectrum showed signal above noise. As the X-ray emission from the host galaxy (such as from X-ray binaries or supernovae remnants) is generally dominant only in the `soft' X-ray band \citep[$E\leq 2\ \rm{keV}$,][]{wang_understanding_2010}, whereas the AGN emission dominates at $E>2\ \rm{keV}$, we can safely assume that there should be no contamination by the host galaxy emission in the range of the spectrum we are fitting.}



\subsection{Properties of {\it the Saguaro}}

We leveraged the available data for {\it the Saguaro} to construct a comprehensive picture of the system: (1) Section~\ref{sec:host} outlines the host galaxy properties; (2) Section~\ref{sec:morphology} examines the morphology; (3) Section~\ref{sec:xrayprops} details its X-ray properties; and (4) Section~\ref{sec:SMBH} derives its $\MBH$ and discusses its implications.

\subsubsection{Host galaxy integrated properties}
\label{sec:host}
\begin{figure}
    \centering
    \includegraphics[width=1.0\linewidth]{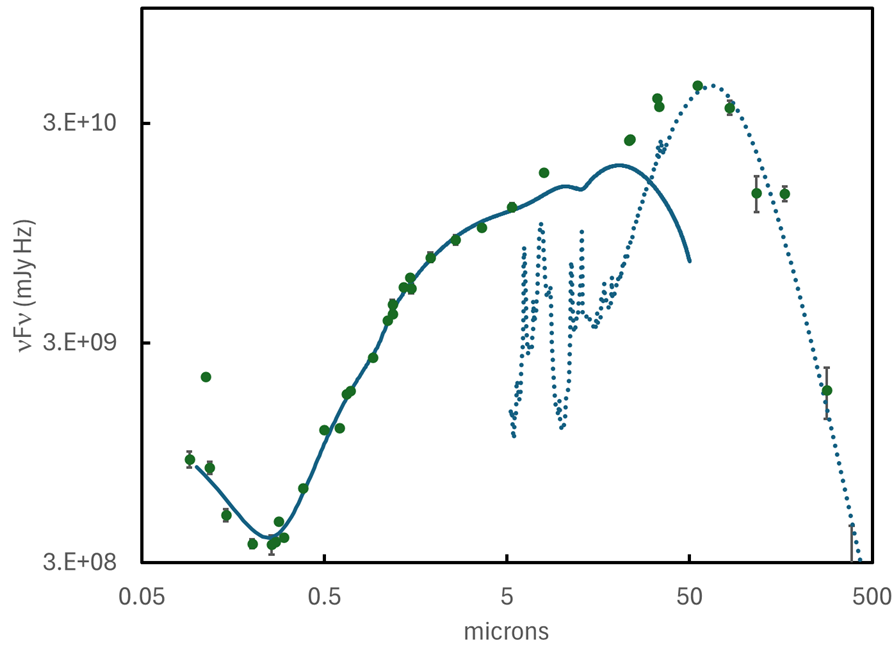}
    \caption{{ The rest-frame SED of {\it the Saguaro}. The template of a $\log_{10}(L_{\rm IR}/L_{\odot}) = 11.5$ star-forming LIRG (\citealt{rieke_determining_2009}; dashed line) is fitted by least squares to the longest-wavelength data points from Table~\ref{tab:data} ($\lambda_{\rm rest} > 80~\mu$m) and subtracted from the total SED to isolate the nuclear component (solid line) after removing the star-forming contribution. This fit assumes that, over the fitted wavelength range, the AGN contribution is much smaller than the star-forming component, as generally observed in typical AGN SEDs (e.g., \citealt{xu_agns_2020, bernhard_post-herschel_2021, lyu_agn_2022}).} 
}
    \label{fig:irebg6sec}
\end{figure}

Figure~\ref{fig:wise} provides an overview of {\it the Saguaro} and the HST and JWST data used to derive its properties. Stellar masses of $\log_{10}(\Mstar/M_{\odot}) \approx 11.3$ have been derived using the FAST code from \citet{kriek_dust_2013} in the recent literature \citep{naidu_hduv_2017, liu_super-deblended_2018, cleri_clear_2023}, often based on photometry compiled by \citet{skelton_3d-hst_2014}. However FAST depends strongly on rest-frame optical colors, which for this galaxy are heavily contaminated by the infrared excess (see Figure~\ref{fig:irebg6sec}). Therefore, to ensure the robustness of the previous measurements, we re-determined the galaxy mass using the AGN-host decomposition (see Section \ref{sec:agn_host_decomposition}) to obtain the intrinsic stellar SED, as illustrated in Figure~\ref{fig:agn-host-dec}, and running {\sc bagpipes} (\citealt{carnall_vandels_2019}) for the synthesis modeling. We find $\log_{10}(\Mstar) = 11.3 \pm 0.1$ (nominal uncertainties only). This confirms the previous values and appears to be inconsistent with  the value of $\log_{10}(\Mstar) \approx 10.46$ from \citet{bongiorno_mbh-m_2014}\footnote{They already reported that the fit for this particular object was challenging (see their Section 4 and Figure 2).}. 

\begin{deluxetable}{lccc}
\tabletypesize{\footnotesize}
\label{agnlist}
\tablecaption{The SED measurements of {\it the Saguaro} compiled from the literature, corresponding to those shown in Figure~\ref{fig:irebg6sec}.} 
\tablewidth{0pt}
\setlength{\tabcolsep}{1.5pt}
\tablehead{
\colhead {rest-frame $\lambda$} & 
\colhead {$\nu F_{\nu}$} &
\colhead {Error}  &
\colhead {Reference}\\
\colhead {($\mu$m)} & 
\colhead {(mJy\,Hz)} &
\colhead {(mJy\,Hz)}  &
\colhead {}
}
\startdata
0.091		&	8.89E+08	&	7.42E+07	&	1	\\
0.111		&	2.12E+09	&	3.61E+07	&	1	\\
0.118		&	8.14E+08	&	5.37E+07	&	2	\\
0.144		&	4.96E+08	&	3.23E+07	&	3	\\
0.201		&	3.68E+08	&	1.70E+07	&	3	\\
0.257		&	3.63E+08	&	3.63E+07	&	3	\\
0.270		&	3.73E+08	&	1.49E+07	&	3	\\
0.282		&	4.64E+08	&	8.50E+06	&	3	\\
0.299		&	3.91E+08	&	3.33E+06	&	3	\\
0.381		&	6.57E+08	&	2.09E+06	&	3	\\
0.498		&	1.21E+09	&	2.60E+06	&	3	\\
0.663		&	1.77E+09	&	2.38E+06	&	3	\\
0.697		&	1.82E+09	&	2.14E+06	&	3	\\
0.919		&	2.58E+09	&	2.17E+06	&	3	\\
1.111		&	3.81E+09	&	1.90E+06	&	3	\\
1.179		&	4.49E+09	&	2.25E+08	&	3	\\
1.181		&	4.08E+09	&	1.58E+06	&	4	\\
1.360		&	5.38E+09	&	2.20E+06	&	3	\\
1.473		&	5.99E+09	&	1.86E+06	&	3	\\
1.492		&	5.32E+09	&	2.66E+08	&	4	\\
1.903		&	7.37E+09	&	3.69E+08	&	4	\\
2.612		&	8.87E+09	&	4.43E+08	&	4	\\
5.321		&	1.25E+10	&	5.61E+08	&	5	\\
7.960		&	1.79E+10	&	5.24E+08	&	5	\\
23.691	&	2.55E+10	&	3.40E+08	&	6	\\
33.960	&	3.57E+10	&	9.23E+08	&	5	\\
54.974	&	4.45E+10	&	1.48E+09	&	5	\\
82.919	&	3.54E+10	&	2.63E+09	&	5	\\
116.106	&	1.45E+10	&	2.69E+09	&	5	\\
165.837	&	1.43E+10	&	1.09E+09	&	5	\\
281.877	&	1.83E+09	&	4.80E+08	&	5	\\
385.669	&	2.73E+08	&	1.68E+08	&	5	\\
\enddata
References: (1) \citet{oesch_dearth_2018-1}; (2) \citet{borys_relationship_2005}; (3) \citet{rieke_jades_2023}, JADES DR3 NIRCam photometry (GOODS-N v1.0; Extension 6) {\tt circ0}, corrected to total flux (factor of 1.25); (4) \citet{teplitz_spitzer_2011}, { All the IRAC measurements were reduced by a factor of 1.22 to bring them into agreement with the JADES photometry at similar wavelengths: this factor was determined by comparing the  measurements at IRAC Band 1 and F356W; comparison of the model fluxes with the other IRAC  bands showed a similar adjustment was needed for them}; (5) \citet{liu_super-deblended_2018}; (6) \citet{hanish_spitzer_2015}.
\label{tab:data}
\end{deluxetable}

Previous works (e.g., \citealt{donley_agn_2010}) suggest that {\it the Saguaro} hosts a deeply embedded AGN. To prove this, we selected the highest quality photometric measurements for this object from the recent literature to generate the SED shown in Figure~\ref{fig:irebg6sec} and listed in Table~\ref{tab:data}.  
We have decomposed the SED into two components: one according to a $\log_{10}(L_{\text{IR}}/L_\odot)=11.5$ LIRG template from \citet{rieke_determining_2009} and the second as the remaining flux. The decomposition is non-degenerate for this case because of the extreme luminosity of the star-forming component, and the selected template is in agreement with the luminosity range found to fit the far infrared of galaxies in this redshift range (\citealt{rujopakarn_mid-infrared_2013}).

{ The strong emission from the very near infrared to 5 microns of the remaining SED (see Figure \ref{fig:irebg6sec}) indicates the presence of a very deeply embedded AGN (i.e., Compton-thick or nearly so, e.g.,  \citealt{kirkpatrick_role_2015, lyu_active_2024})}. We will find that its luminosity approximates the Eddington limit for the embedded SMBH (Section~\ref{sec:SMBH}), indicating that little luminosity escapes.  \citet{donley_agn_2010} show that the mid infrared spectrum is a virtually featureless power law rising rapidly toward longer wavelengths. Although some H$\alpha$ escapes, the hydrogen lines are heavily obscured, as shown by H$\beta$ not being detected. The source may be analogous to extreme objects like IRAS05189-2524 and IRAS08572+3915 (e.g., \citealt{severgnini_x-ray_2001, efstathiou_herschel_2014}), deeply embedded AGNs with luminosities approaching $10^{12}$ L$_\odot$ but still letting sufficient light through to show AGN emission lines. 

We can determine a star formation rate (SFR) from Figure~\ref{fig:irebg6sec}. The total infrared luminosity from the star-forming template is $3.2 \times 10^{12}$ L$_\odot$, corresponding to an integrated SFR of 500 M$_\odot$/yr, consistent with the result from \citet{bongiorno_mbh-m_2014} but somewhat lower than some of the previous estimates (e.g., \citealt{del_moro_mid-infrared_2016, cleri_clear_2023}). The specific SFR (sSFR) is then $\approx10^{-8.60}\,\rm yr^{-1}$, which is exceptionally high for its redshift, placing it along the upper envelope of the Main Sequence of star-forming galaxies (\citealt{speagle_highly_2014, popesso_main_2023}). The luminosity of the AGN component is then $4.6 \times 10^{12}$ L$_\odot$, assuming essentially all of it emerges in the infrared.

\subsubsection{System morphology }
\label{sec:morphology}

Figure~\ref{fig:wise} shows images of {\it the Saguaro} across the HST and NIRCam wavelength range. In the rest-frame UV, the emission is dominated by a compact central source surrounded by a faint halo, consistent with a face-on spiral host. As we move into the rest-frame optical, the extended galaxy becomes increasingly prominent (\citealt{donley_agn_2010}).  At redder wavelengths, the central core increasingly dominates the emission, as expected from the SED in Figure~\ref{fig:irebg6sec}. This effect is visible across the individual bands (see Figure~\ref{fig:wise}) and also in our AGN–host decomposition discussed later.

\subsubsection{X-ray properties}
\label{sec:xrayprops}

To fit the X-ray spectrum, we chose a simple phenomenological model and a slightly more complex theoretical model. The phenomenological model is composed of a power-law, a neutral absorber at the redshift of the source, and a local neutral absorber to correct for the Galactic absorption. Due to the redshift of the source and the limited amount of counts, our spectrum is limited to $E>3.5\,\rm{keV}$, making it challenging to fit at the same time both the power-law slope and the source obscuration; therefore, we fixed the photon index to a canonical $\Gamma=1.8$. The spectrum is well fitted by our simple model, and neither additional soft emission or reflection components improve our fit significantly. Our fit provides an obscuration of $N_H = (3.4^{+1.0}_{-0.8}) \times 10^{23}\,\mathrm{cm}^{-2}$ and an intrinsic (i.e. absorption-corrected) 2-10 keV (rest-frame) luminosity of $L_{2\text{–}10\,\mathrm{keV}} = (7.9^{+2.1}_{-1.6}) \times 10^{43}\,\mathrm{erg\,s^{-1}}$.

We tried to physically model the spectrum via the MYTorus model \citep{murphy_x-ray_2009}, which self-consistently accounts for photoelectric absorption, Compton scattering, and fluorescent line emission in a toroidal geometry. MYTorus assumes a uniformly dense, toroidal obscurer with a fixed half-opening angle of $60^{\circ}$. This model can be used in both “coupled” and “decoupled” modes \citep{yaqoob_nature_2012}, the latter of which mimics a patchy/clumpy absorber, making it well-suited for studying moderately to heavily obscured sources. We utilized MYTorus in its “coupled” configuration and due to the low number of counts, certain parameters were fixed during fitting. Namely the slope at $\Gamma=1.8$, and the scaling factors of the scattered continuum and line emission were both fixed at 1. We found results consistent with those from the phenomenological modeling, and obtained a torus obscuration along the line-of-sight of $N_{H,\mathrm{los}} \gtrsim 2.6 \times 10^{23}\,\mathrm{cm}^{-2}$, an inclination angle of $\geq 60^\circ$, and an intrinsic luminosity of $L_{2\text{–}10\,\mathrm{keV}} = (7.5^{+1.3}_{-1.1}) \times 10^{43}\,\mathrm{erg\,s^{-1}}$. All in all, our measurements are consistent with the previous literature { (i.e., \citealt{donley_agn_2010, bongiorno_mbh-m_2014, del_moro_mid-infrared_2016})}. 


From our estimate of the intrinsic X-ray luminosity and using the infrared luminosity of the AGN component as $L_{bol}$, we estimate the bolometric correction factor ($k_{\text{bol}} \equiv L_{\text{bol}}/L_{\text{X, 2–10 keV, int}}$) to be $\approx$ 200.  For an AGN of the appropriate luminosity, the expected value is $\approx$ 50  (\citealt{duras_universal_2020, auge_accretion_2023}). This suggests that the intrinsic X-ray luminosity has been underestimated by as much as a factor of four, or that the source is modestly X-ray weak.

As a check on this result,  we have derived $L_{\text{bol}}$ from [Ne\,{\sc v}]$\lambda3426$, which can be used as a tracer of $L_{\text{bol}}$ in obscured AGNs (e.g., \citealt{gilli_x-ray_2010, barchiesi_cosmos2020_2024}). We adopted the calibration presented in \citet{feuillet_star_2024}\footnote{Adapted from the calibration proposed in \citet{satyapal_discovery_2007} and rescaled to  [Ne\,{v}]$\lambda3426$ as discussed in \citet{feuillet_star_2024}.} (see their Section 4), where $(L_{\rm bol}/ \text{erg}\,\text{s}^{-1}) = 0.94\,\log_{10} (L_{[\rm Ne\,{v}]\lambda3426} /\rm  erg\,s^{-1})+6.51$. We model the [Ne\,{\sc v}]$\lambda3426$ line from NIRSpec/PRISM data with both {\sc MSAEXP} and our internal pipeline { (already used in \citealt{rinaldi_deciphering_2025, rinaldi_not_2025})}, obtaining a consistent flux of $(9.03 \pm 0.71) \times 10^{-17}\,\rm erg\,s^{-1}\,cm^{-2}$, corresponding to $L_{\rm bol} \approx (2.56 \pm 0.22) \times 10^{46}\,\rm erg\,s^{-1}$, where the uncertainty is driven primarily by the signal-to-noise of the line. The correlation itself is established at $\approx 4.5 \sigma$ \citep{barchiesi_cosmos2020_2024}, so the final estimate is $(6 \pm 1.5) \times 10^{12}$ L$_\odot$, consistent  with our result from integrating the infrared SED associated with the AGN \footnote{{ In contrast, \citet{cleri_clear_2023} measured a [Ne\,{\sc v}]$\lambda3426$ flux from HST/WFC3 G102 slitless spectroscopy that is nearly twice as large as our value. This difference is expected from the different spectral normalizations adopted. In this work, we rescale the spectrum to the model photometry of the central source, providing a measurement tied to the nuclear component. By contrast, the slitless measurement is tied to the total extracted light, for which the continuum normalization can include host-galaxy contribution. We therefore adopt our nuclear-normalized [Ne\,{\sc v}] flux.}}. This lends additional support to attributing this source component to an embedded AGN.

To further assess the X-ray nature of this source, we examine the relation between the rest-frame UV and X-ray monochromatic luminosities, $L_{2500\,\text{\AA}}$ and $L_{2\,\mathrm{keV}}$, through the optical-to-X-ray index ($\alpha_{\rm OX}$; \citealt{lusso_tight_2016}). Without applying any dust correction to $L_{2500\,\text{\AA}}$, we derive $\alpha_{\rm OX} \approx -1.25$. { However, the interpretation of this value is not straightforward. The $\alpha_{\rm OX}$ relation was calibrated primarily for unobscured AGNs, where the rest-frame UV emission is expected to be dominated by the accretion disk. This assumption may not hold for the Saguaro, since the observed rest-frame UV/optical light may include a contribution from the host galaxy and/or compact nuclear star formation.}

{ At the same time, the source exhibits a high $k_{\rm bol}$ ($\approx200$), suggesting that it may be X-ray faint relative to its inferred bolometric output. This apparent discrepancy may arise from residual X-ray absorption not fully accounted for by the applied column density, uncertainties in the inferred bolometric luminosity, contamination of $L_{2500\,\text{\AA}}$, or an intrinsically weak X-ray corona. Reconciling the observed $\alpha_{\rm OX}$ with a substantially more X-ray-weak regime would require additional attenuation, corresponding to $A_V \approx 2$~mag for a Calzetti reddening law \citep{calzetti_dust_2000}. While this is slightly higher than the upper limit of $A_V$ inferred from our {\sc bagpipes} SED modeling, the values remain broadly consistent within the uncertainties and potential systematics in the SED decomposition.}

{ The detection of broad H$\alpha$ emission provides an additional, although not definitive, constraint on the obscuring geometry. It suggests that at least part of the broad-line region remains visible along our line of sight. However, our X-ray modeling still indicates substantial obscuration, consistent with an inclined torus-like geometry. Therefore, broad H$\alpha$ does not rule out residual X-ray obscuration. A natural interpretation is a partially obscured or patchy configuration, in which the broad-line region is not fully hidden while the more compact X-ray-emitting corona remains attenuated. H$\beta$ cannot currently provide an independent test, as it is blended in the PRISM spectrum and is not covered by the available higher-resolution data. We therefore treat $\alpha_{\rm OX}$ as an indicative, but not decisive, diagnostic for this source. A similar optical/X-ray mismatch was recently reported for the heavily obscured SMG ALESS073.1 at $z=4.76$, where a very broad H$\alpha$ component reveals the BLR despite a high X-ray column density close to the Compton-thick regime \citep{parlanti_ga-nifs_2024}.}

\subsubsection{Properties of the SMBH}
\label{sec:SMBH}
{\it The Saguaro} is covered by NIRSpec G235H/F170LP observations, which we reduced using {\sc MSAEXP}. Its spectrum reveals a broad H$\alpha$ component (see Figure \ref{fig:wise}) with $F(\text{H}\alpha) = (1.03 \pm 0.05) \times 10^{-16}~\mathrm{erg\,s^{-1}\,cm^{-2}}$ and a FWHM of $2521 \pm 256$ km\,s$^{-1}$. We also detect the [N\,{\sc ii}]$\lambda\lambda6548, 6583$ doublet with a flux ratio of $3.01 \pm 0.52$, in agreement, within uncertainties, with the theoretical value of 3.05 \citep{dojcinovic_flux_2023}. 
The broad H$\alpha$ was also found in previous works by \citealt{Swinbank_rest-frame_2004}, \citet{bongiorno_mbh-m_2014}, and \citealt{Wirth_team_2015}.

{ Because of the uncertainties affecting the broad H$\alpha$ measurement, including possible biases from residual cosmic rays and detector artifacts in the high-resolution data, the lack of high-resolution H$\beta$ coverage needed to correct for dust attenuation, and the difficulty of isolating the intrinsic AGN continuum, we do not adopt an H$\alpha$-based black-hole mass as our fiducial estimate. Instead, we estimate the black-hole mass from the X-ray luminosity.\footnote{{ Nonetheless, adopting the relation from \citet{reines_dwarf_2013}, we obtain a lower limit on $\MBH$ that is consistent with both literature estimates and our X-ray-based inferred value.}}}  We follow the calibration from \citet{LaMassa_estimating_2025}. However, we need to correct for the very large $k_{\rm bol}$ (200)  compared with the relation assumed there (\citealt{Lusso_x-ray_2010}). The result is $\log_{10}(\MBH/M_\odot) = 8.12$, with errors that are difficult to estimate because of the complexities in determining the intrinsic X-ray properties. This is consistent with the prior estimate of $\log_{10}(\MBH/M_\odot) = 7.99 \pm 0.3$ by \citet{bongiorno_mbh-m_2014}, but is higher because of our allowance for the large $k_{\rm bol}$.  
Thus, the resulting Eddington Limit is $4.2 \times 10^{12}$ $L_\odot$, essentially the same as our value for the AGN luminosity, i.e., the source is accreting close to the Eddington rate. 
The nucleus is  broadly along the local $\MBH\text{--}\Mstar$ relation \citep{greene_megamaser_2016} with $\MBH/\Mstar \approx$ 0.0004, i.e. within the scatter in that relation and the uncertainties in $\MBH$ and $\Mstar$.  

\subsection{Resemblance to LRDs}

In this section, we emphasize the similarity between {\it the Saguaro} and canonical LRDs by focusing on two defining features: its characteristic “V-shaped” SED (Section \ref{sec:prism_vshape}), which we show originates from the central region (Sections \ref{sec:agn_host_decomposition}$\text{--}$\ref{sec:nuclear_v-shape}), { the presence of a dense gas absorber in He\,{\sc i}$\lambda10830$ (Section \ref{HeI_saguaro})}, and the unresolved, point-like morphology it would present if observed at high redshift (Section \ref{sec:redshift_exp}). { Moreover, we show that its sub-mm and X-ray detections, together with the presence of [N\,{\sc ii}]$\lambda6548,6583$, do not contradict its resemblance to LRDs (Sections~\ref{sec:submm}, \ref{sec:xraydet}, and \ref{NII_saguaro}).
}

\subsubsection{The V-shape of the Saguaro}\label{sec:prism_vshape}

The NIRSpec/PRISM spectrum of {\it the Saguaro} provides one of the clearest examples of a “V-shaped” spectrum, emerging at a rest-frame wavelength comparable to those observed in canonical LRDs. We measure UV and optical slopes of $-0.812$ and $0.901$ (see Figure \ref{fig:wise}), respectively. We find the spectrum is fully consistent with the definition of canonical LRDs proposed by \citet{kocevski_rise_2024}. We show these slopes in Figure \ref{fig:wise}. { By adopting the same methodology as \citet{setton_little_2024}, we find a spectral break at $3668^{+5}_{-28}\,\text{\AA}$, very close to the Balmer limit ($3646\,\text{\AA}$) as inferred by \citet{setton_little_2024} in their sample of LRDs.} Interestingly, the “V-shaped” spectrum is observed in NIRSpec/PRISM data that sample nearly the central region of the galaxy, with the shutter positioned close to its nucleus.

\subsubsection{Multiwavelength AGN-host image decomposition}\label{sec:agn_host_decomposition}

We further validate the ``V-shaped" spectrum of \textit{the Saguaro} using multiband photometric decomposition to separate the AGN from its host galaxy. We employ the \texttt{GALFITM} package \citep{hausler_megamorph_2013}, an extension of the widely used \texttt{GALFIT} software \citep{peng_detailed_2002, peng_detailed_2010}, to simultaneously model our NIRCam and HST images, allowing parameter variations as functions of wavelength. This method effectively leverages multiband data, producing more robust fits compared to single-band decompositions, while accounting for morphological wavelength variations. \texttt{GALFITM} has been widely utilized for structural decompositions of AGN-host galaxy systems at cosmic noon \citep[e.g.,][]{zhuang_evolutionary_2023, zhuang_archival_2023, gillman_midis_2025}.

We model the AGN as a point source and the host galaxy with a single Sérsic profile to capture its dominant structure ({ see Figure \ref{fig:residual} in Appendix}). While the host exhibits spiral arms, they are far enough from the galactic center that they do not contaminate the AGN flux significantly. Our Sérsic model thus represents the azimuthally averaged light profile of the disk. A more complex decomposition, incorporating a galactic bar and a nearby stellar clump, located 0.18$''$ southwest of the nucleus, is presented later in this section ({ see Figure \ref{fig:residual_complex} in Appendix}). However, we adopt the simpler AGN–galaxy decomposition as our fiducial result, since it captures the dominant components while avoiding parameter degeneracy associated with more complex models. We do not include a bulge component, as neither the NIRCam nor HST images reveal a prominent bulge component. If present, the bulge is likely a pseudo-bulge: classical bulges typically have effective radii larger than 1 kpc \citep{dimauro_structural_2019}, which would be resolved by NIRCam. For example, the FWHM of the NIRCam/F200W PSF is 0.064$''$, corresponding to 0.55 kpc at $z=2$, smaller than the typical radius of a classical bulge. Pseudo-bulges generally have bulge-to-total flux ratios below 0.2 \citep{fisher_structure_2008}, and since {\it the Saguaro} is brighter than its host in most bands, any contribution from a pseudo-bulge is likely minor.

In the {\tt GALFITM} configuration, we use Chebyshev polynomials to model the wavelength dependence of structural parameters, allowing for smooth variation across filters while mitigating overfitting. The magnitudes of both the AGN and Sérsic components are left free in all bands. The Sérsic index and half-light radius  are allowed to vary quadratically with wavelength, following the approach of \citet{hausler_megamorph_2013} and \citet{zhuang_archival_2023}, which provides sufficient flexibility to capture real structural changes while maintaining stability in low signal-to-noise filters. To avoid introducing artificial wavelength-dependent distortions, we fix the ellipticity and position angle across all bands. We also assume the AGN and host galaxy share a common center, fixed in all bands. Nearby galaxies are manually masked in the decomposition.

\begin{figure}
    \centering
    \includegraphics[width=1\linewidth]{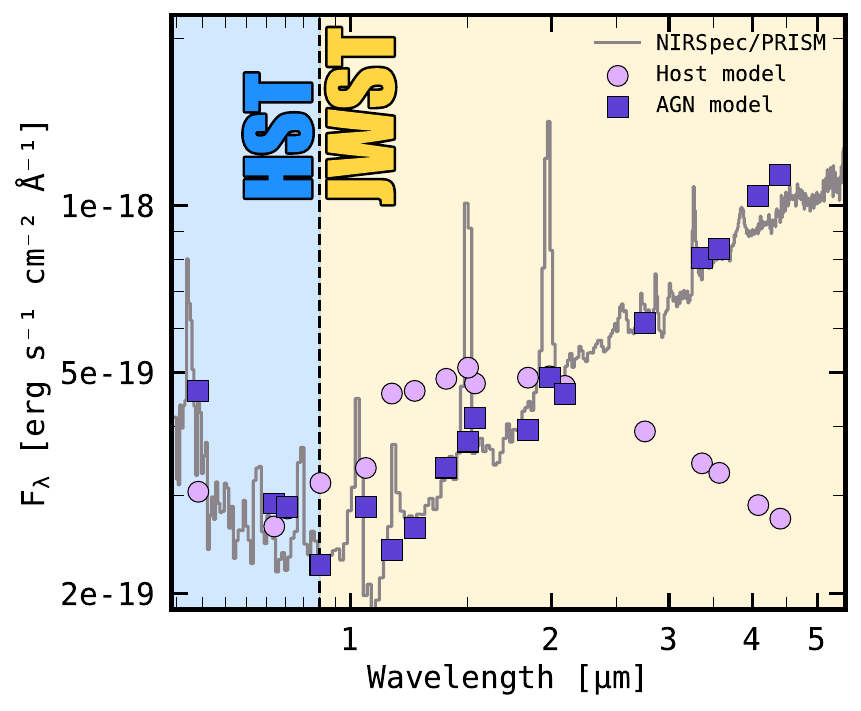}
    \caption{The NIRSpec/PRISM spectrum of {\it the Saguaro} { at observed wavelengths} (using the standard reduction from the DAWN JWST Archive), shown alongside the AGN–host decomposition photometry from {\sc GALFITM}, scaled to match the NIRSpec/PRISM flux level { at 2~$\mu$m}. The vertical line indicates the transition from HST to JWST coverage, below which the UV photometry is provided solely by HST.}
    \label{fig:agn-host-dec}
\end{figure}

We present our results in Figure~\ref{fig:agn-host-dec}, which shows the modeled photometry for both the host galaxy and the AGN, each normalized to match the NIRSpec/PRISM spectrum { at $2~\mu$m}, along with the spectrum itself. The AGN–host decomposition reveals that the characteristic “V-shaped” SED is well reproduced by the point-like source used to model the AGN component, distinct from the SED of the host galaxy. { As a test, we also estimated the relative AGN and host contributions as a function of radius (see Figure \ref{fig:flux_fraction_components} in Appendix). We find that the AGN component dominates at small radii, while the host contribution becomes dominant at larger radii.} { If one considers the standard LRD criteria (see \citealt{rinaldi_way_2026} for an overview), Saguaro would not be robustly selected as an LRD. First, classical color cuts based on fixed observed-frame filters (see e.g. \citealt{barro_extremely_2024, labbe_population_2023}), are not ideally suited to Saguaro because, at its redshift, they do not probe the useful rest-frame wavelength ranges. A more appropriate alternative is the criterion proposed by \citet{kocevski_rise_2024}, which adopts a bandpass-shifted approach. By exploring different apertures, from 0.1\arcsec\; to 0.6\arcsec, we find that only the smallest aperture, i.e., 0.1\arcsec, satisfies the color selection. More critically, Saguaro fails the compactness requirement, since the source is resolved in F444W (when host and AGN are not separated). This is also why it was not included in the recent, more inclusive census of LRDs by \citet{rinaldi_way_2026}. These results show that lower-$z$ LRD analogs with prominent host emission may be missed by standard selections, and that identifying such systems may require separating the compact nuclear component from the extended host contribution.} A similar conclusions has been also suggested recently by \citet{billand_investigating_2025}. 

The host galaxy properties are consistent with normal disk galaxies. The half-light radius of the host galaxy is 5.8 kpc at rest-frame 0.15 $\mu$m and decreases to 3.7 kpc at 1.5 $\mu$m, consistent with the known trend of decreasing size with wavelength (e.g., \citealt{vulcani_galaxy_2014}). The measured Sérsic index remains near $n\approx1$ across all bands, typical of disk galaxies. The SED of the host galaxy, as shown in Figure~\ref{fig:agn-host-dec}, resembles that of normal spiral galaxies in the local universe \citep{brown_atlas_2014}, showing a Balmer break and a downturn beyond $\approx2$ $\mu$m (rest-frame 0.7 $\mu$m), in contrast to the steeply rising AGN SED. In the rest-frame UV, the $F_\lambda$ decreases with wavelength, but the slope is flatter than that of the AGN.

Interestingly, residuals near the galaxy center reveal asymmetric, off-center features located $\approx0.1$\arcsec ($\approx0.8$\,kpc) from the nucleus. This structure is unlikely to be a bulge given its offset from the nucleus. It resembles the off-center blobs seen in LRDs by \citet{chen_physical_2025}, which they suggest might be merging companion galaxies, disturbed host structures, or AGN-illuminated nebular emission. Notably, {\it the Saguaro} has two neighboring galaxies to the east and southeast at projected distances of 24 and 26 kpc. Their JADES photometric redshifts from the JADE catalog are 2.04 and 2.03, respectively, consistent with the spectroscopic redshift of {\it the Saguaro} ($z=2.0145$), suggesting potential physical association. If spectroscopically confirmed, they could possibly be evidence of ongoing minor mergers. Alternatively, the asymmetric, off-center features could be also a stellar clump near the nucleus, which is common seen in $z\approx2$ galaxies (e.g., \citealt{kalita_rest-frame_2024}).

Finally, we tested a sophisticated model including a bar and a stellar clump in the modeling, fitted alongside the AGN and disk. The clump, located 0.18$''$ southwest of the nucleus, is clearly seen in short-wavelength images. The bar, oriented northeast–southwest, is most prominent at long wavelengths, likely due to  relatively older stellar populations common in bars. The bar is modeled with a Sérsic profile with a center tied to the disk. The fluxes across multiple bands are allowed free, while the Sérsic and radius are allowed to vary quadratically with wavelengths. The stellar clump is modeled as a point source. We find an improvement of residuals, and the AGN SED retains its V-shape. However, the added complexity introduces degeneracies and unphysical wiggles. For example, in the F200W band, where the bar is not visible, the model increases bar flux to account for the asymmetric, extended feature mentioned above, and thus reduces the inferred AGN flux. Moreover, at long-wavelengths, the stellar clump is blended with the AGN due to lower resolution. To avoid such artifacts, we adopt the simpler AGN–Sérsic decomposition as our baseline. The sophisticated model confirms that contamination from the substructures is $<10$\%, so it does not affect our main conclusions.

Although the AGN–host decomposition reveals the spatial coincidence of the AGN with the observed “V-shaped” SED feature, it remains unclear whether it arises solely from AGN activity or is also influenced by an additional component, such as hot, young stars in a circumnuclear starburst ring. These structures are commonly observed in the central regions of active galaxies at low redshift ($z \lesssim 0.1$). Although they have typical radii $\gtrsim 0.5$ kpc (e.g., \citealt{kotilainen_near-infrared_2000, bohn_goals-jwst_2023}), which would be resolved with NIRCam at the redshifts probed here ($z\approx2$), it remains uncertain whether similar structures exist on smaller, unresolved spatial scales. In this regard, ALMA observations have revealed a more complex picture in the nuclear regions of high-$z$ galaxies, often showing clumpy, compact star formation and gas distributions (e.g., \citealt{gilli_alma_2014, hodge_alma_2019}), { which may cause contamination to our AGN-only SED at rest-frame UV and blue-optical SED, i.e. roughly $\approx0.12\text{--}0.4\,\mu$m (e.g., \citealt{gonzalez_delgado_role_1999}).}


\subsubsection{The “V-shape” lives at the center of the galaxy: insights from NIRSpec/PRISM}\label{sec:nuclear_v-shape}

The multiwavelength AGN–host image decomposition revealed that the “V-shaped” SED arises primarily from the AGN component used to model the central source. To investigate this detail, we further analyzed the NIRSpec/PRISM observations, which cover the central region of the object (see Figure \ref{fig:wise}).   In particular, we performed an {\it ad hoc} reduction of the NIRSpec/PRISM data using {\sc MSAEXP}. However, instead of the default nodding technique (adopted by the DAWN JWST Archive), which can suppress extended emission in the outskirts, we applied a customized background subtraction method following the procedure described in \citet{de_graaff_rubies_2025}, now fully implemented in {\sc MSAEXP}. In this approach, a global master sky spectrum is subtracted rather than using image differencing, effectively avoiding self-subtraction and preserving any extended structure. This strategy is essential to determine whether the “V-shaped” feature is truly concentrated in the nuclear region of {\it the Saguaro}, as it maximizes the recovered flux across the entire spatial profile.

By preserving the full spatial structure in the 2D spectrum and avoiding the background suppression due to default nodding techniques, we were able to extract one-dimensional spectra at each spatial pixel along the cross-dispersion axis, including those just beyond the nuclear regions of the galaxy. As described in the previous section, the AGN–host decomposition performed with {\sc GALFITM} provided photometry for both the compact central AGN and the more extended host component. Therefore, for each spatial row, we compare the extracted spectrum with the corresponding AGN–host decomposed photometry, normalized { at 2 $\mu$m}. 

\begin{figure*}
    \centering
    \includegraphics[width=1.0\linewidth]{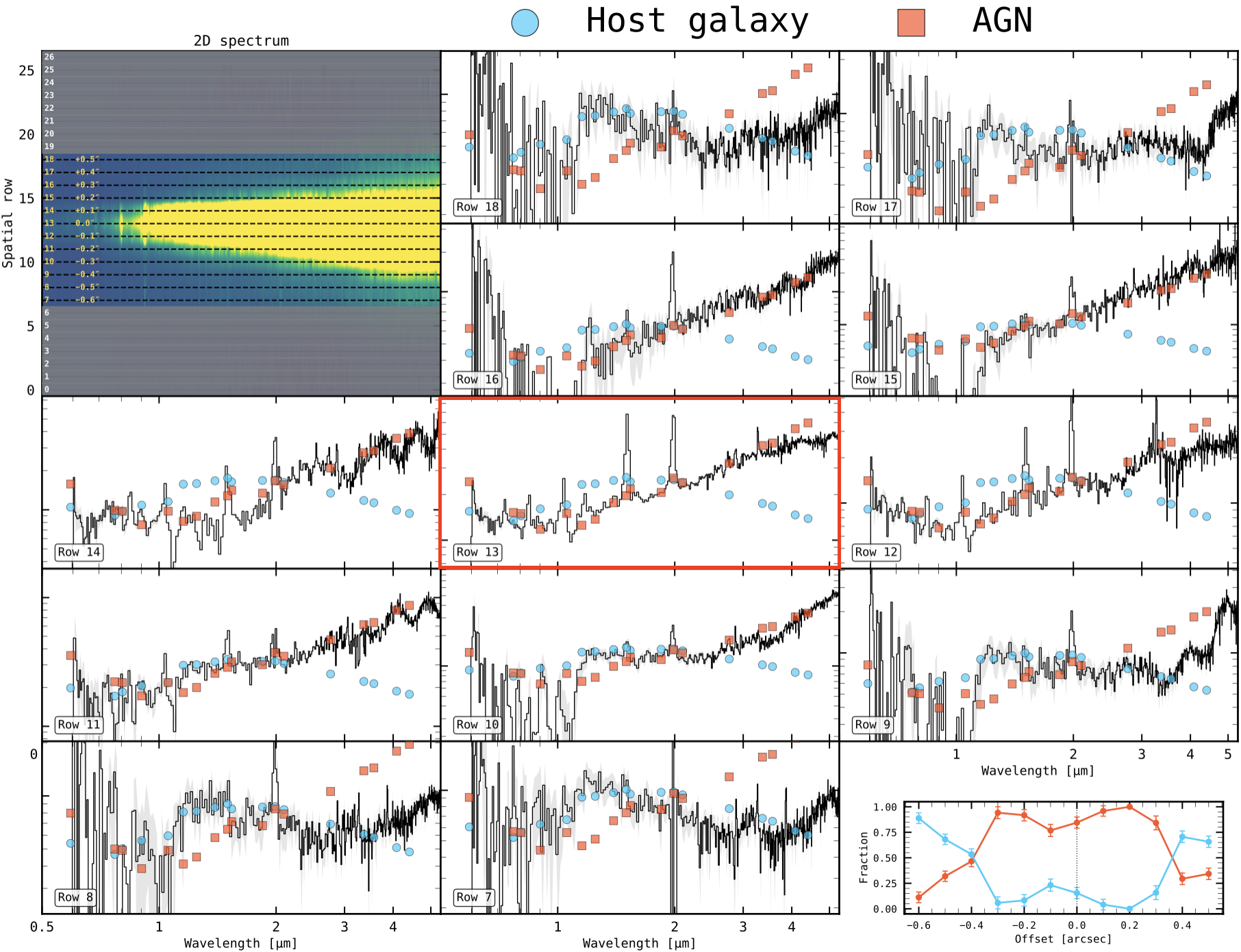}
    \caption{Extraction of row-by-row 1D spectra from the NIRSpec/PRISM 2D data of {\it the Saguaro}. { Spectra are shown in $F_\lambda$ (y-axis) vs. $\lambda$  (x-axis); spectra are shown in observed frame.} Each subplot on the right shows the extracted spectrum (black) for a given spatial row, along with the corresponding host galaxy and AGN photometry (from {\sc GALFITM}) normalized { (at $2\,\mu$m)} to match the extracted spectrum at each row. The center of the trace is located at row 13 { (highlighted in red)}. We also report the distance from the center of the spectral trace. { We fit the spectrum extracted in each row as a linear combination of the AGN and host components inferred from the {\sc GALFITM} decomposition. This allows us to estimate the relative AGN and host contributions as a function of distance from the center. We find that the “V-shaped” component dominates in the central row, but becomes diluted at larger radii, where the host contribution takes over.}}
    \label{fig:spectral_traces}
\end{figure*}

We show the extracted spectra in Figure~\ref{fig:spectral_traces}. While this analysis pushes the NIRSpec/PRISM data to its limits, it clearly reveals a strong contrast between the central “V-shaped” SED and the surrounding regions. The characteristic “V-shape” is most prominent within the central $\approx0.1\text{--}0.2$\arcsec (corresponding to $\approx0.8\text{--}1.5$ kpc at $z = 2.0145$). { To explore this,  we fit the spectrum extracted in each row as a linear combination of the AGN and host components derived from the GALFITM decomposition. This provides a quantitative estimate of the relative AGN and host contributions as a function of distance from the center. We find that the “V-shaped” component dominates in the central row, but is rapidly diluted at larger radii, where the host emission becomes dominant.} 

{ Although the NIRSpec/PRISM spectrum provides spatial information along the slit direction, the row-by-row extractions should not be interpreted as independent measurements at the native 0.1\arcsec\ pixel scale. The effective spatial resolution is set by the NIRSpec PSF, which redistributes nuclear light into adjacent rows. Therefore, the transition observed in the extracted spectra should be regarded as PSF-smoothed rather than spatially sharp. With this caveat, we still find a coherent spatial trend. The central rows show the strongest “V-shaped” continuum, while at offsets of order $\approx0.1\text{--}0.2$\arcsec\ and beyond this component becomes progressively diluted and the spectrum increasingly resembles the host-dominated component inferred from the independent {\sc GALFITM} decomposition. This supports the conclusion that the “V-shaped” SED is mainly associated with the innermost region of the galaxy.}

{ These findings also underscore the importance of slit placement when attempting to identify such features from spectroscopy alone. The presence and detectability of a “V-shaped” SED in a spectroscopic extraction could strongly depend on whether the aperture or shutter covers the compact nuclear component. This is particularly relevant when the imaging data do not have sufficient S/N or resolution to enable a reliable image decomposition. In such cases, slight slit or shutter misalignments could dilute or miss the nuclear component, potentially hiding “V-shaped” cores within otherwise normal-looking galaxies.}

\subsubsection{The presence of dense gas absorber in {\it the Saguaro}}\label{HeI_saguaro}
\begin{figure}
    \centering
    \includegraphics[width=1\linewidth]{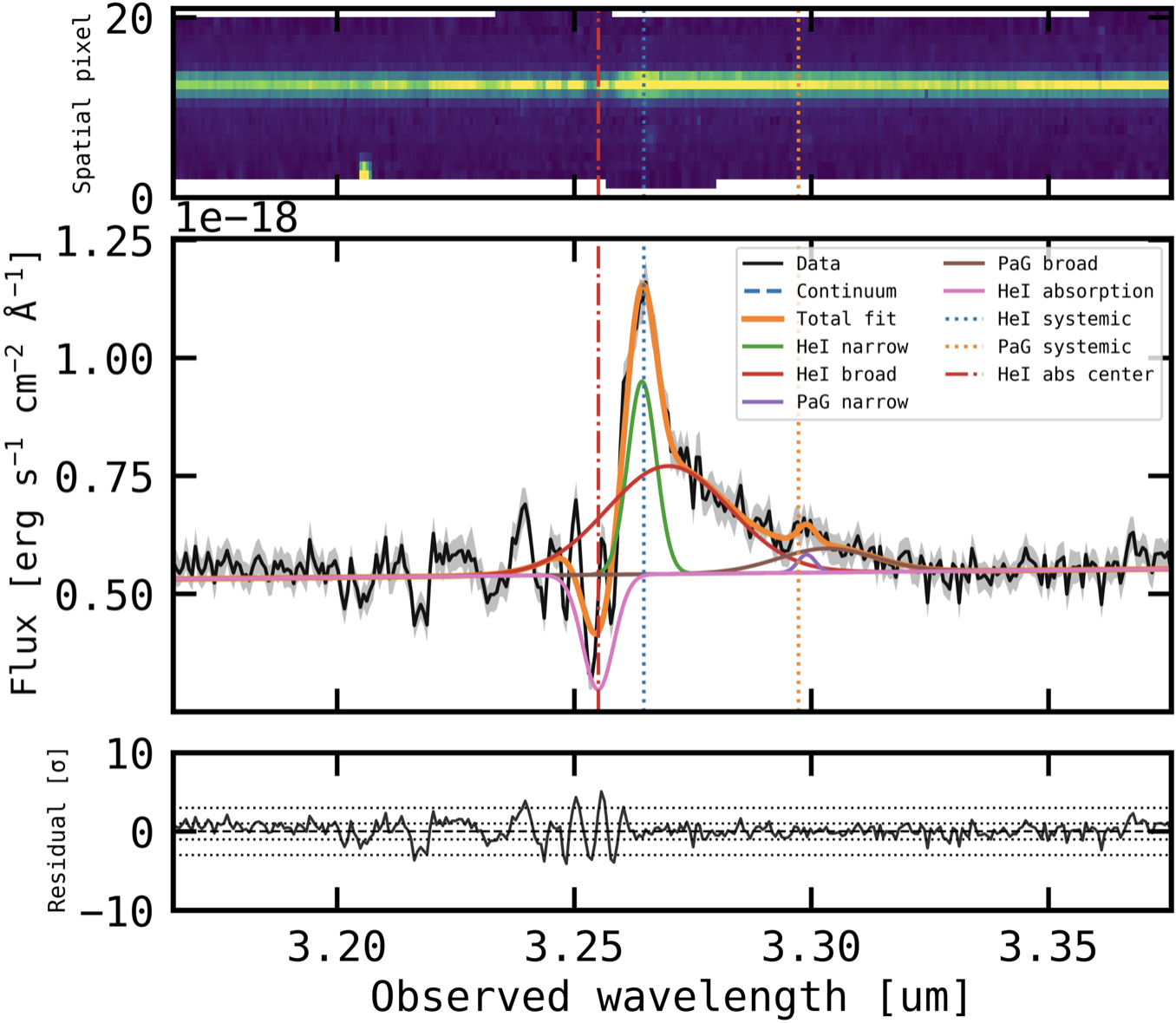}
    \caption{{ G395H/F290LP spectrum covering the He\,{\sc i}$\lambda10830$ and Pa-$\gamma$ emission-line complex. A clear absorption feature is visible in both the 2D and 1D spectra, indicating the presence of dense neutral or partially ionized gas along the line of sight, likely located in the nuclear region close to the accretion disk and/or the broad-line region.}
}
    \label{fig:HeI_abs}
\end{figure}

{ Recent models suggest that the peculiar spectra of LRDs cannot be explained by a standard broad-line region alone. Instead, they may be shaped by radiative-transfer effects through dense, partially ionized, high-column-density gas around the central source, producing their distinct features, including Balmer absorption \citep[e.g.,][]{inayoshi_extremely_2025, ji_blackthunder_2025, naidu_black_2025, de_graaff_remarkable_2025}. Nonetheless, a similar absorption feature can also appear in He\,{\sc i}$\lambda10830$, which provides a complementary tracer of dense, partially ionized gas because it arises from the metastable $2^3S$ level and does not require the extreme population of the short-lived H$(n=2)$ level needed for Balmer absorption.

Therefore, we searched for similar absorption features in the high-resolution spectra covering {\it the Saguaro}. We stress that this inspection is strongly limited by the relatively poor data quality, and defer a more detailed analysis to a future paper based on higher-quality NIRSpec/IFU data from PID 10311. Nonetheless, this search remains informative, as it allows us to assess possible similarities and differences with respect to what is commonly observed in canonical LRDs.

First, we do not identify any clear absorption feature in H$\alpha$ (see Figure \ref{fig:wise}). Whether this reflects the intrinsic absence of Balmer absorption or is simply due to the limited data quality remains unclear. However, in the G395H/F290LP spectrum, both the 2D and 1D data show evidence for an absorption feature associated with the He\,{\sc i}$\lambda10830$ line (see Figure \ref{fig:HeI_abs}). This feature appears as a strongly blueshifted absorption component superimposed on the overall emission-line complex, together with broad Pa-$\gamma$ emission.

Our fit strongly favors a model including the absorption component. Specifically, we find $\Delta_{\rm BIC}>100$ relative to a model without absorption, corresponding to an $\approx5\sigma$ detection. Following the methodology outlined in \citet{loiacono_big_2025}, we infer a velocity offset of $\Delta v=-885.2\pm36.5$~km~s$^{-1}$ and a FWHM of $634.3\pm84.2$~km~s$^{-1}$. The corresponding column density of helium in the metastable $2^3S$ level is $N({\rm He\,{\sc I}}\,2^3S)\approx(1.73-2.50)\times10^{13}$~cm$^{-2}$. 

Although H$\alpha$ absorption is commonly observed in canonical LRDs, the situation observed in {\it the Saguaro} (namely, the absence of clear H$\alpha$ absorption together with the detection of He\,{\sc i}$\lambda10830$ absorption) is not unique. A similar combination has also been reported in J1047 (\citealt{lin_discovery_2025}; local LRD analog) and GLIMPSE-17775 (\citealt{kokorev_deepest_2025}; a $z=3.501$ LRD), showing that this combination is present in other claimed LRDs or LRD analogs. 
}

\subsubsection{The submillimeter flux}\label{sec:submm}

The current strong submm flux seen in {\it the Saguaro} is clearly in violation of the properties of canonical LRDs (\citealt{casey_upper_2025}). However, as shown in Figure~\ref{fig:irebg6sec}, this is completely driven by the very high rate of star formation, which is likely to deplete the  interstellar gas and fade quickly ($\lesssim$ 20 Myr)  (e.g., \citealt{forster_schreiber_nature_2003}) with an upper limit of $20\text{--}60$ Myr set by the gas depletion timescale (\citealt{liu_dust_2024}).  This could leave a dust enshrouded AGN that emits weakly in the submm (unless it has successfully blown away its cocoon). In any case, ULIRGs and high Eddington ratio AGNs are not strongly correlated (\citealt{rieke_low_2025}), so this source component does not seriously undermine the analogy with LRDs.

\subsubsection{The X-ray detection}\label{sec:xraydet}
{ Recent studies have shown that LRDs are notoriously faint in the X-ray regime. In most cases, X-ray data do not yield individual detections, and stacking analyses have only provided upper limits (e.g., \citealt{ananna_x-ray_2024, yue_stacking_2024}). This has led to the interpretation that LRDs are, on average, intrinsically X-ray weak, heavily obscured, or both. In this context, the X-ray luminosity inferred for {\it the Saguaro} ($\log_{10}\left(L_{2\text{--}10\,\mathrm{keV}}/\mathrm{erg\,s^{-1}}\right)=43.88\pm0.07$) lies at the upper end of the existing upper limits (spanning $\log_{10}(L_{2\text{--}10\,\mathrm{keV}}/\mathrm{erg\,s^{-1}})\approx42.5$--$44.5$; see Figure 1 in \citealt{yue_stacking_2024}). A few confirmed X-ray-detected LRDs are currently known, though. \citet{kocevski_rise_2024} reported two cases: PRIMER-COS 3866 at $z=4.66$, with $\log_{10}(L_{2\text{--}10\,\mathrm{keV}}/\mathrm{erg\,s^{-1}})=44.7\pm0.2$, and JADES 21925 at $z=3.1$, with $\log_{10}(L_{2\text{--}10\,\mathrm{keV}}/\mathrm{erg\,s^{-1}})=43.73\pm0.06$. The former is substantially more X-ray luminous than {\it the Saguaro}, while the latter is comparable. More recently, \citet{hviding_x-ray_2026} reported another prominent X-ray-detected LRD (called XRD), with an X-ray luminosity comparable to PRIMER-COS 3866 but at lower redshift ($z\approx3.28$). 

Therefore, although the X-ray detection of {\it the Saguaro} may appear exceptional at first glance, it instead places the source among the rare X-ray-detected LRDs currently known. Moreover, our analysis suggests that {\it the Saguaro} is both X-ray weak and strongly obscured, as indicated, for example, by its high bolometric correction ($k_{\rm bol}\approx200$). We therefore conclude that its X-ray detection does not contradict its resemblance to the LRD population. Rather, as discussed for XRD (\citealt{hviding_x-ray_2026}), {\it the Saguaro} may represent a transitional object, further strengthening its role as a key system for investigating the eventual fate of LRDs at later cosmic times.}

\subsubsection{The presence of [N\,{\sc ii}]$\lambda6548,6583$ in the Saguaro}\label{NII_saguaro}
{ Another apparent difference between the Saguaro and classical LRDs at higher redshift is the clear presence of [N\,{\sc ii}]$\lambda6548,6583$, which has not been commonly reported in classical high-$z$ LRDs. One could argue that this feature weakens the comparison between {\it the Saguaro} and the high-redshift LRD population. However, recent DESI results (\citealt{collaboration_data_2026}) have revealed LRD-like systems at $z<1$, where both weak and strong [N\,{\sc ii}]$\lambda6548,6583$ emission has been reported (see \citealt{ding_discovery_2026, park_new_2026, lin_lrds2_2026}). In particular, \citet{ding_discovery_2026} reported five LRD-like systems at $z\approx0.2\text{--}0.4$ whose spectral shapes closely resemble those of classical high-$z$ LRDs. Similar to {\it the Saguaro}, these systems also show relatively strong [N\,{\sc ii}] emission, with line ratios comparable to that measured in {\it the Saguaro} ($\log_{10}({\rm [N\,II]}/\mathrm{H}\alpha)\approx0.05\pm0.05$).
The presence of [N\,{\sc ii}] therefore does not necessarily rule out a connection with the classical LRDs. Instead, it may indicate that {\it the Saguaro}, and other systems with comparable [N\,{\sc ii}]/H$\alpha$ ratios, are more chemically evolved than classical high-$z$ LRDs, which are likely hosted by low-metallicity systems. This could reflect their more prominent host-galaxy contribution and a more advanced evolutionary stage. Consistently, both {\it the Saguaro} and the DESI LRD-like systems from \citet{ding_discovery_2026} lie on the local $\MBH\text{--}M_{\bigstar}$ relation, suggesting that they may represent later evolutionary stages of compact, rapidly accreting systems observed at much higher redshifts, in which the host galaxy has caught up with the growth of the central supermassive black hole. Finally, visual inspection of the recently reported low-$z$ LRDs from DESI (\citealt{park_new_2026, ding_discovery_2026, lin_lrds2_2026}) shows that these systems are often embedded within extended galaxies. This further suggests that some apparent contradictions may arise from the stronger contribution of the host galaxy in these systems compared to their classical high-$z$ counterparts.} 

\subsubsection{How we wonder what you would have looked like in the early Universe}\label{sec:redshift_exp}

We showed that {\it the Saguaro} exhibits a “V-shaped” spectrum originating from its nuclear region, with the “V-break” occurring at the characteristic wavelength reported in the literature (\citealt{setton_little_2024}). We then demonstrated that AGN–host image decomposition confirms the spatial coincidence of this spectral shape with the central AGN { as well as the presence of a dense gas absorber in {\it the Saguaro}}. Finally, we highlighted that its sub-mm flux, X-ray detection{ , and the presence of [N\,{\sc ii}]$\lambda\lambda6548,6583$} do not violate its resemblance to the broader LRD population. This strong resemblance to LRDs naturally raises the question of whether this object represents a low-redshift analog of LRDs. However, some key differences must be taken into account. First, compactness is a defining characteristic of LRDs, whereas our object (considering both AGN and host) is clearly extended. Second, {\it the Saguaro} is observed at $z \approx 2$, while canonical LRDs are typically observed at $z \gtrsim 4$–5. Additionally, its face-on disk morphology and far-infrared properties differ significantly from what is currently known about LRDs (although the latter may be a transient phase). { Furthermore, the prominent [N\,{\sc ii}] emission in {\it the Saguaro} may indicate a host galaxy that is more chemically enriched and massive than those typically associated with canonical LRDs.} These differences could reflect evolutionary changes over cosmic time. 

This ambiguity makes  a compelling case for investigating whether it can be considered as a possible analog of canonical LRDs at later epochs. Therefore, we investigate how {\it the Saguaro} would appear if observed (unchanged in its intrinsic properties) at higher redshifts.
We emphasize that this is a simplified, toy experiment. 
Nevertheless, this exercise raises important questions about whether current observations of LRDs at high redshift are systematically biased in recovering host galaxy properties, as recently suggested by \citet{pacucci_cosmic_2025} and \citet{billand_investigating_2025}. In this regard, \citet{rinaldi_not_2025} has already shown that approximately 30\% of photometrically selected high-redshift LRDs exhibit complex or disturbed morphologies { (which they interpret as a proxy for merger activity in at least a subset of LRDs)}, underscoring the need to carefully consider observational limitations, such as image depth, which might bias this fraction.

To investigate how cosmological surface brightness dimming \citep{tolman_estimation_1930} affects the detectability of extended emission in {\it the Saguaro} at high redshifts,  we performed pixel-by-pixel SED fitting using {\sc EazyPy} \citep{brammer_eazy_2008}. { The redshift was fixed to that of the galaxy derived from the NIRSpec/PRISM spectrum.} We adopted  the {\tt BLUE\_SFHZ\_13} template set along with an AGN+torus template, which includes redshift-dependent star formation histories and dust attenuation. { In particular, the AGN+torus template is based on the NIRSpec/PRISM spectrum of the $z=4.50$ source from GO-1433 in the MACS0647 field (PI: D. Coe; see \citealt{killi_deciphering_2023}). This source exhibits a spectral shape representative of LRDs, and therefore it is useful to describe the nuclear region of {\it the Saguaro}.} The fit was restricted to NIRCam bands only, taking advantage of their superior spatial resolution. All bands were PSF-matched to F444W using \textsc{STPSF} (\citealt{perrin_updated_2014}), and SEDs were extracted from each pixel after masking nearby contaminants. 

Each pixel’s SED was then redshifted to the target redshift, and the galaxy’s appearance at earlier cosmic times was reconstructed following a \textsc{FERENGI}-like methodology \citep{barden_ferengi_2008}. This includes angular size rescaling using the proper angular diameter distance ratio, full $(1+z)^{-4}$ surface-brightness dimming, and bandpass shifting. An evolutionary brightening term was also applied (see next paragraph). The redshifted SEDs were projected onto a blank canvas, convolved with the corresponding PSF, and injected into a real background extracted from a region close to the source. The background was cleaned by masking sources and replacing contaminated regions with nearby noise samples to preserve realistic structure and depth.

We stress that for this toy experiment, our goal is purely to visualize how cosmological surface-brightness dimming affects the extended host galaxy. To that end, we adopt the phenomenological prescription implemented in \textsc{FERENGI}, applying a uniform brightening of $M_{\mathrm{evo}}(z) = x\,z + M_0$ with $x = -1$ \citep{barden_ferengi_2008}. Following the discussion in \citealt{barden_ferengi_2008}, this choice is made purely for visual purposes, to illustrate the impact of surface-brightness dimming with redshift. We note that a physically realistic treatment would require evolving the AGN and host components separately, with the AGN scaled according to the evolution of the quasar bolometric luminosity function \citep[e.g.,][]{shen_bolometric_2020, zhang_trinity_2024}, and the host following the stellar mass-to-light evolution observed in star-forming galaxies \citep[e.g.,][]{treu_massive_2004}.

\begin{figure*}
    \centering
    \includegraphics[width=1.0\linewidth]{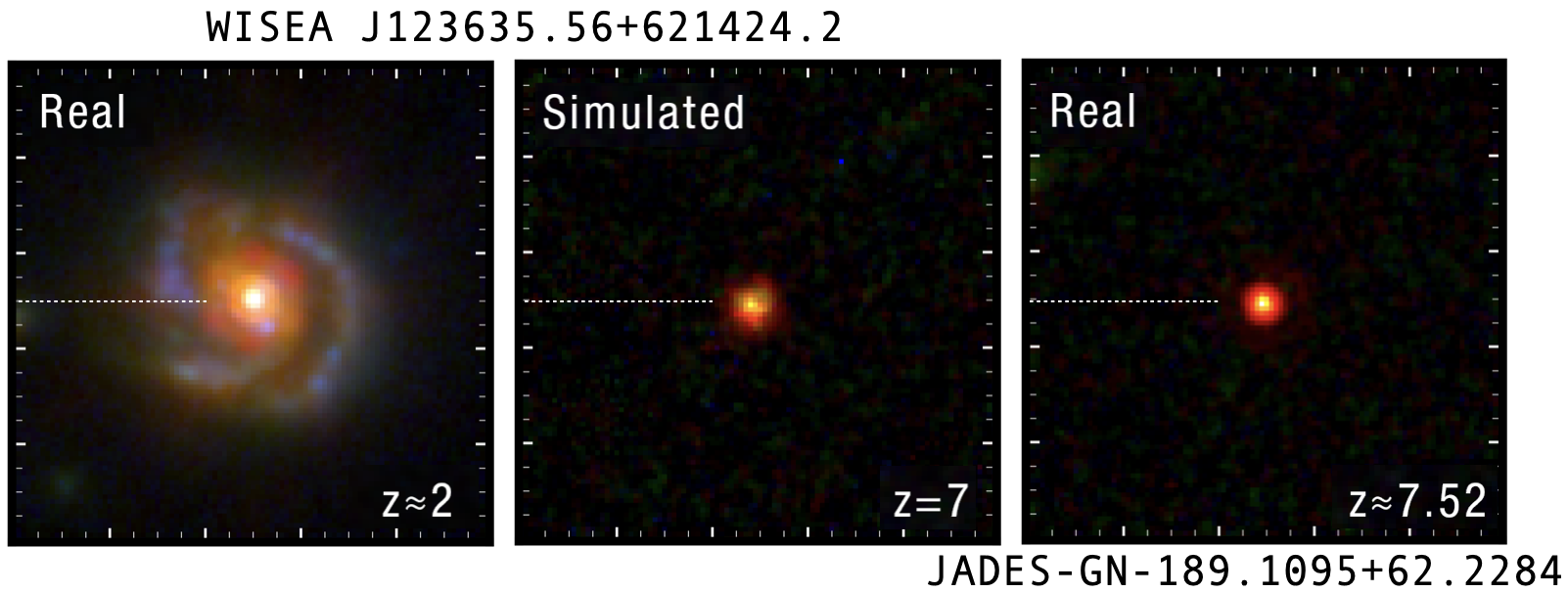}
    \caption{Visual demonstration of cosmological surface-brightness dimming and its impact on the host galaxy. { Left:} the real {\it Saguaro} (WISEA~J123635.56+621424.2) observed at $z = 2.0145$. { Center:} the same galaxy artificially redshifted to $z = 7$ using the \textsc{FERENGI} prescription \citep{barden_ferengi_2008}. { Right:} A representative LRD at comparable redshift from \citet{rinaldi_not_2025}. All cutouts are $3\arcsec \times 3\arcsec$. The toy experiment demonstrates how the host galaxy becomes increasingly suppressed in its outskirts, while the central regions remain detectable, mirroring the effect observed in LRDs. { For these RGB images, we make use of the code {\sc Trilogy} (D. Coe) with some customized steps. Briefly, each band is independently scaled using a sigma-clipped estimate of the background and a logarithmic stretch. This scaling is designed to suppress noise-dominated pixels (and thus equalize the background across all images), preserve faint extended emission, and avoid excessive saturation of the brightest regions. The RGBs are based on F444W, F277W, and F090W.}
}
    \label{fig:simulated_LRD}
\end{figure*}

We show the results of this toy experiment in Figure~\ref{fig:simulated_LRD}, which showcases three RGB images: the real object (on the left), its simulated appearance at $z=7$ (at the center), and, for comparison, a representative LRD at similar redshift (on the right). This experiment demonstrates how cosmological surface-brightness dimming preferentially suppresses low-surface-brightness emission in galaxy outskirts while leaving compact central components detectable. 
This differential effect, which is consistent with the findings of \citet{calvi_effect_2014}, highlights how extended stellar light can easily fall below detection thresholds at high redshift, potentially explaining the lack of extended host galaxies around  LRDs as redshift increases.

\section{How much we never see: Are LRDs embedded in extended hosts?}

\subsection{The stacked UV profile of LRDs}
We showed that {\it the Saguaro} (1) exhibits a “V-shaped” SED at its center, as revealed by both the AGN–host decomposition and the spatially resolved NIRSpec/PRISM spectrum, and (2) that if redshifted to high redshift (unchanged in its properties), the cosmological surface brightness dimming would preferentially suppress flux from the outskirts, leaving only the central AGN  visible. Given the striking resemblance between {\it the Saguaro} and LRDs, and taking into account the intrinsic size evolution of galaxies with redshift (e.g., \citealt{ormerod_epochs_2024, genin_dawn_2025}), this naturally raises the question of whether observational biases significantly affect our current view of canonical LRDs at high redshift.

To explore this possibility, we analyzed the sample of 99 photometrically selected LRDs at $z \approx 4\text{--}8$ presented in \citet{rinaldi_not_2025}, noting that not all of them necessarily host an AGN\footnote{However, diagnostic diagrams based on emission lines likely indicate a mixed origin, involving both star formation and AGN activity (\citealt{rinaldi_not_2025}).}. We stacked all sources using filters probing the rest-frame UV range of $1500\text{--}2000$~{\AA} (F090W, F115W, and F150W), after PSF-matching each image with \textsc{STPSF} to the reddest available filter in that wavelength range (F150W). Median-stacked cutouts were centered on the LRD positions, enhancing coherent low-surface-brightness features while suppressing outliers. To ensure that the stack captures only emission associated with the LRDs, all sources beyond 0.3\arcsec\, from the center were masked out prior to stacking. This method enables the isolation of extended UV emission.

\begin{figure*}
    \centering
    \includegraphics[width=0.49\linewidth]{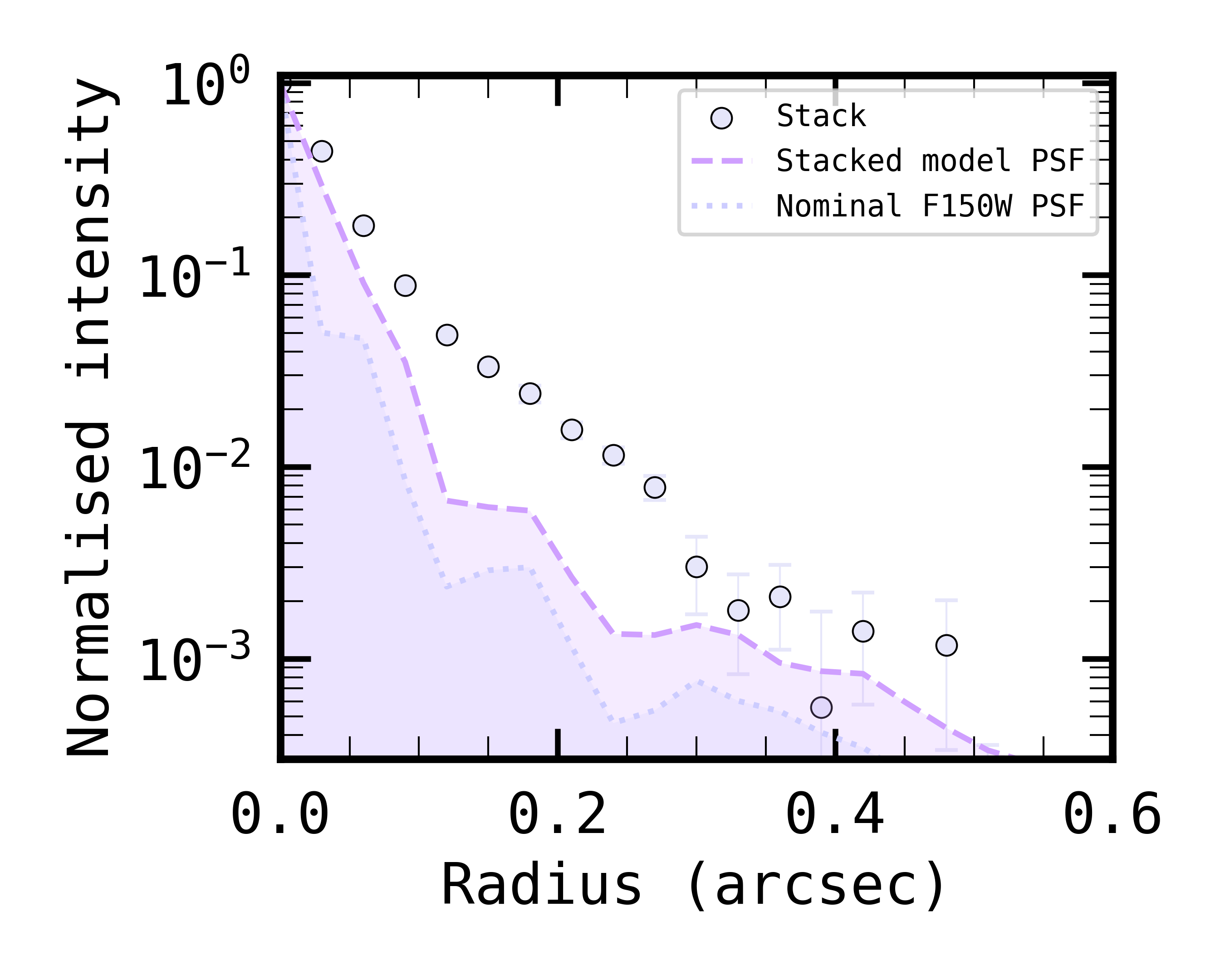}
    \includegraphics[width=0.49\linewidth]{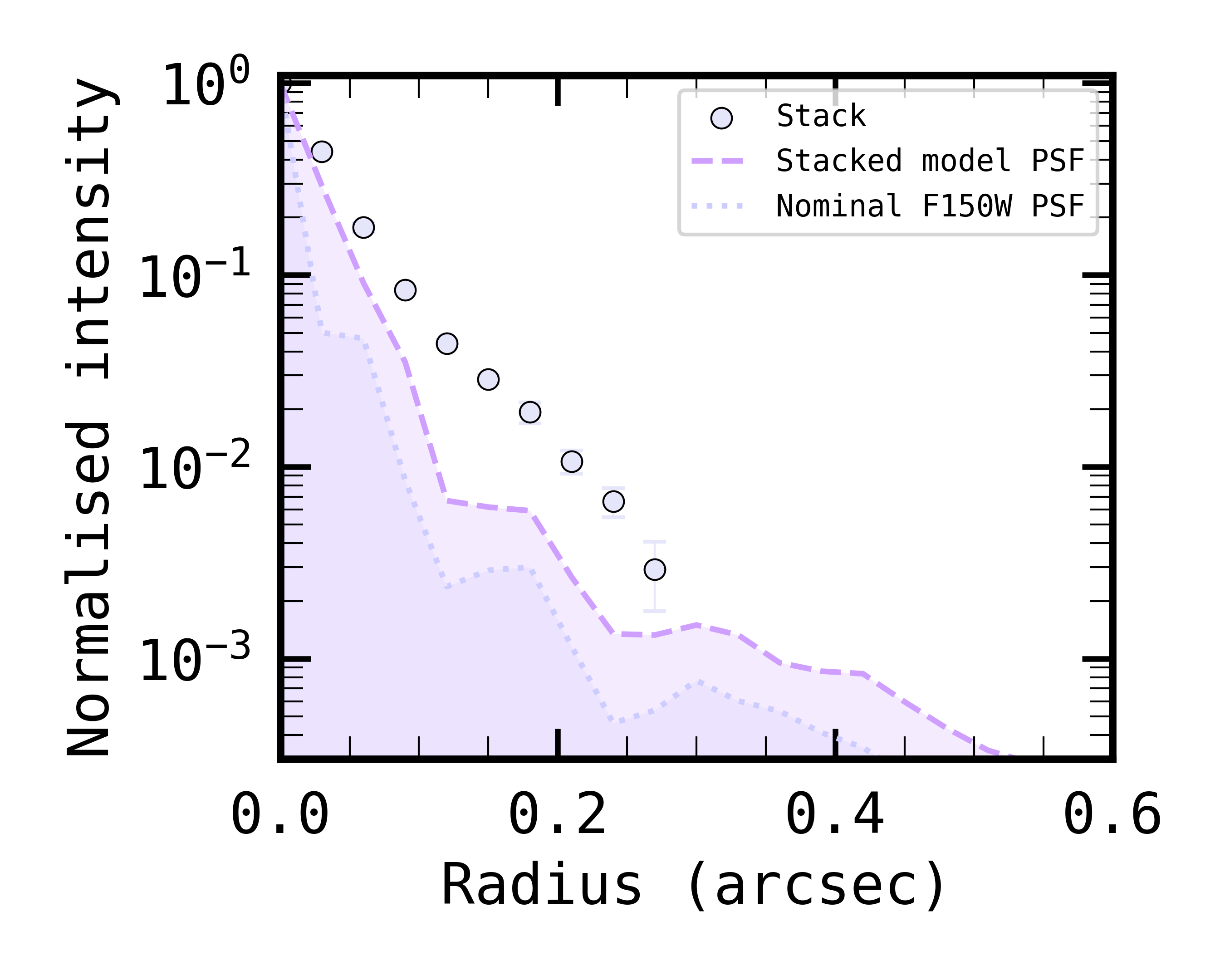}
    \caption{{ Left:} Radial profile of the stacked image of 99 photometrically selected LRDs from \citet{rinaldi_not_2025} in the rest-frame UV ($1500$–$2000$~{\AA}).
{ Right:} Same as on the left, but with the background artificially elevated to the highest level that does not cause the surrounding region to become strongly negative. To emphasize the extended nature of the stacked UV profile of the LRDs, we note that our simulated stacked PSF was constructed using a spatial offset of 1 pixel ($\approx30$ mas for the NIRCam short wavelength channel), which is significantly larger ($\approx 6\times$) than the typical internal astrometric scatter in the JADES images, thus representing an upper limit on the observed stacked PSF size. Altogether, this experiment reveals that the radial profile of the stacked LRDs declines less steeply than expected, suggesting that an extended structure may lie hidden beneath the observational noise.
}
    \label{fig:psf_experiment}
\end{figure*}

We then derived the radial surface brightness profile of the stacked image and compared it to two PSF references: (1) the nominal PSF used for matching (F150W), and (2) a simulated stacked PSF created by repeating our full stacking procedure on synthetic NIRCam PSFs\footnote{{ To do so, we made use of 99 simulated NIRCam PSFs.}}. To ensure realism, we applied random V3-angle rotations and spatial offsets before stacking, mimicking the astrometric and orientation variations present in the real dataset. { We randomly applied a 1-pixel shift along both the $x$ and $y$ directions, corresponding to 30 mas for the NIRCam short-wavelength channel.} { No measurable systematic dependence is found on the V3 angles, while the results are strongly sensitive to the applied offset. In particular, the 30 mas shifted PSF is broader than the nominal PSF by a factor of $\approx2\text{--}3$ at $r\approx0.05$\arcsec, increasing to $\approx3\text{--}5$ at $r\approx0.1$\arcsec$\text{--}0.3$\arcsec, as one can see in Figure \ref{fig:psf_experiment}. In our case, the applied offset is deliberately exaggerated, being approximately six times larger than the typical internal astrometric scatter in the JADES images ($\lesssim5$ mas), and was adopted to estimate a conservative upper limit on the impact of residual astrometric uncertainties on the inferred extended host UV emission associated with LRDs.}

Figure~\ref{fig:psf_experiment} (on the left) shows the result assuming a conservative background subtraction, where the outermost profile points are not allowed to average negative. Figure~\ref{fig:psf_experiment} (on the right) applies a more aggressive subtraction\footnote{{ All JADES images are already background-subtracted; however, small residuals may still persist. We therefore applied an additional background subtraction after stacking.}}, allowing for negative values in the outskirts. In both cases, significant excess emission is seen beyond the PSF, extending to a radius of $\approx0.2$\arcsec, or $\approx2\text{--}3$ kpc in diameter. The agreement between the two profiles down to a normalized intensity of $10^{-2}$ supports the robustness of this detection. This is also consistent with the recent findings of \citet{lin_spectroscopic_2024}, who analyzed a sample of broad H$\alpha$ emitters at $z \approx 4-5$. However, their sources were not explicitly selected as LRDs and were drawn from a significantly smaller sample, yet the results are similar.

Interestingly, we repeated the same stacking analysis using only the subset of LRDs with spectroscopic redshifts (26 galaxies in total). Although this subset represents a modest fraction (approximately 26\%) of the full sample, the results remain unchanged, even when applying aggressive background subtraction { (see Figure \ref{fig:psf_experiment_spec} in Appendix)}. We also repeated the experiment excluding the galaxies classified as having “complex UV morphology” in \citet{rinaldi_not_2025}, and found consistent results, further supporting the idea that canonical LRDs may reside within extended, low-surface-brightness envelopes. { A similar result has also been reported for the optical emission; indeed, \citet{zhang_unveiling_2025} found evidence for a host component through a stacking analysis of more than 200 LRDs in COSMOS-Web, further supporting the presence of a faint extended host emission in LRDs.}

\begin{figure}
    \centering
    \includegraphics[width=1\linewidth]{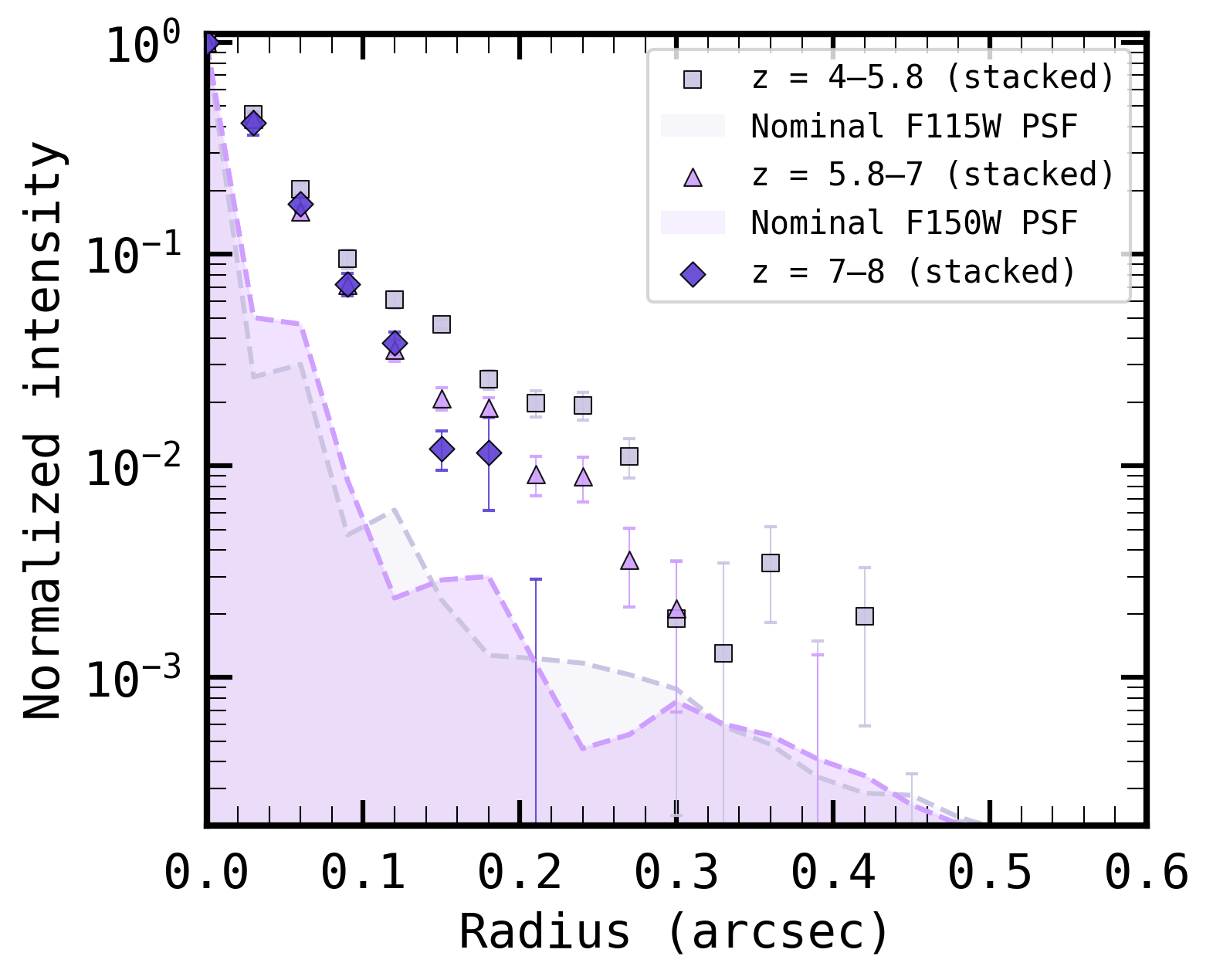}

    \caption{Radial profiles of the stacked rest-frame UV ($1500$–$2000$~\AA) images of photometrically selected LRDs from \citet{rinaldi_not_2025}, shown in redshift bins. For comparison, the nominal PSFs are overplotted: F115W for $z \approx 4\text{--}5.8$ and F150W for $z \approx 5.8\text{--}8$. A slight increase in radial extent is observed when moving from the highest to the lowest redshift bin.}
    \label{fig:stack_bin}
\end{figure}

Moreover, when examining the stacked images in successive redshift bins\footnote{The bins were selected to ensure sufficient statistics in each, accounting for the redshift evolution in the number density of LRDs \citep{kocevski_rise_2024, kokorev_census_2024}.} ($z \approx 4\text{--}5.8$, $z \approx 5.8\text{--}7$, $z \approx 7\text{--}8$), we observe a slight evolution in the extent of the radial profiles from $z \approx 7\text{--}8$ to $z \approx 4\text{--}5.8$, consistent with the expected size evolution of galaxies across this range \citep[e.g.,][]{ormerod_epochs_2024, genin_dawn_2025}. We show this in Figure \ref{fig:stack_bin}. Interestingly, we also find that objects below $z\approx5$ fall almost entirely in the sub-sample of LRDs showing complex UV morphology. The number of such LRDs drops significantly from $z \approx 5$ to $z \approx 8$, with only two objects above $z > 7$ still showing clear morphological complexity. This naturally raises the question of whether this is a true effect, where as soon as the Universe matures, these objects start building more extended envelopes or whether this is a selection effect (\citealt{billand_investigating_2025}). In the latter case, given the small sample size, we caution that the results may be biased, and a larger sample will be required to robustly confirm these findings.

Although the extended emission revealed in the stacked images is smaller than the $\approx4\text{--}6$ kpc hosts identified at $z \approx 2$, it is broadly consistent with the expected size evolution of galaxies. According to \citet{genin_dawn_2025}, disk sizes roughly double between $z \approx 5$ and $z \approx 2$, implying that the compact high-redshift structures we observe could naturally evolve into the more extended hosts seen around systems like {\it the Saguaro} at later times. This strengthens the case for a physical connection between canonical LRDs and their potential lower-$z$ analogs.

\subsection{The LRD visibility with redshift}


To further investigate the idea that our view of LRDs at high redshift may be biased, we develop a simple analytic model of surface brightness dimming to quantify how much of a LRD's extended host galaxy becomes undetectable at progressively higher redshifts. We assume the host follows a circular exponential light profile with a central surface brightness $I_0$ and scale radius $R_s$:

\begin{equation}
I(R) = I_0 \, \exp\left(-\frac{R}{R_s}\right) \, ,
\end{equation}
\noindent
where $I(R)$ is the intrinsic surface brightness at galactocentric radius $R$. The total integrated luminosity of the host is then $L_{\mathrm{host}} = 2\pi I_0 R_s^2$.

Due to cosmological surface brightness dimming, the observed surface brightness is reduced by a factor of $(1 + z)^4$ \citep{tolman_estimation_1930}. We define the maximum detectable radius $R_{\max}$ at which the observed SB equals the detection limit $I_{\mathrm{lim}}$:

\begin{equation}
R_{\max}(z) = R_s \ln\left(\frac{I_0}{I_{\mathrm{lim}}(1+z)^4}\right) \, .
\end{equation}
\noindent
This expression requires  the argument of the logarithm is larger than 1, which implies $I_0 > I_{\mathrm{lim}}(1+z)^4$. Within $R_{\max}$, the fraction of the total host light that remains detectable is computed as:

\begin{equation}
f_{\mathrm{det}}(z) = 1 - \left(1 + \frac{R_{\max}(z)}{R_s}\right) \exp\left(-\frac{R_{\max}(z)}{R_s}\right) \, .
\end{equation}
\noindent
The fraction lost is then $f_{\mathrm{lost}}(z) = 1 - f_{\mathrm{det}}(z)$.

To test this simple model, we adopt the following reference values, motivated by observed LRD properties and JWST detection thresholds:
$I_0 = 100 \, L_\odot\,\mathrm{pc}^{-2}$ (central surface brightness of the host), $R_s = 1.2$ kpc (scale radius, e.g. $\approx$ 0.2\arcsec\, at z = 7), and $I_{\mathrm{lim}} = 0.0079 \, L_\odot\,\mathrm{pc}^{-2}$ (corresponding to $\mu \approx 28.5$ mag/arcsec$^2$ in the rest-UV).
The typical value for $I_0$ was taken from, e.g., \cite{tacchella_sinszc-sinf_2015}, for $z = 2$.
We compute the lost light fraction for redshifts $z = 2$, $5$, $7$, and $9$. The results are summarized in Table \ref{tab:SB_dimming}. If we also account for the observed redshift evolution of galaxy sizes (e.g., $R_s \propto (1+z)^{-m}$; \citealt{ormerod_epochs_2024, genin_dawn_2025}), the maximum detectable radius $R_{\max}$ would shrink significantly in physical units, further compounding the loss of extended light at high redshift.

\begin{deluxetable}{cccc}
\tablecaption{Surface Brightness Dimming Effects\label{tab:SB_dimming}}
\tablehead{
\colhead{$z$} & \colhead{$R_{\max}$ [kpc]} & \colhead{$f_{\mathrm{det}}$} & \colhead{$f_{\mathrm{lost}}$}
}
\startdata
2 & 5.59 & 0.914 & 0.086 \\
5 & 2.73 & 0.664 & 0.336 \\
7 & 1.35 & 0.311 & 0.689 \\
9 & 0.28 & 0.024 & 0.976 \\
\enddata
 
\end{deluxetable}

These results show that, even for relatively compact and moderately bright host galaxies, surface brightness dimming can significantly limit the detection of extended components at high redshift. By $z \gtrsim 7$, more than two-thirds of the host light is lost to cosmological dimming, consistent with the observed compact morphologies of LRDs in JWST imaging.
This calculation agrees with the model by \citet{pacucci_cosmic_2025}, which proposed that, at $z \gtrsim 8$, the observability of LRDs is significantly suppressed by the cosmological surface brightness dimming. Our results are also consistent with the recent findings of \citet{billand_investigating_2025}.

The surface brightness dimming for high redshift LRDs is strong enough that we cannot tell if they are truly isolated point-like sources or if they are the nuclei of more extended galaxies. 

\section{Low-redshift analogs: how typical is {\it the Saguaro}?}


\begin{figure*}
    \centering
    \includegraphics[width=1.0\linewidth]{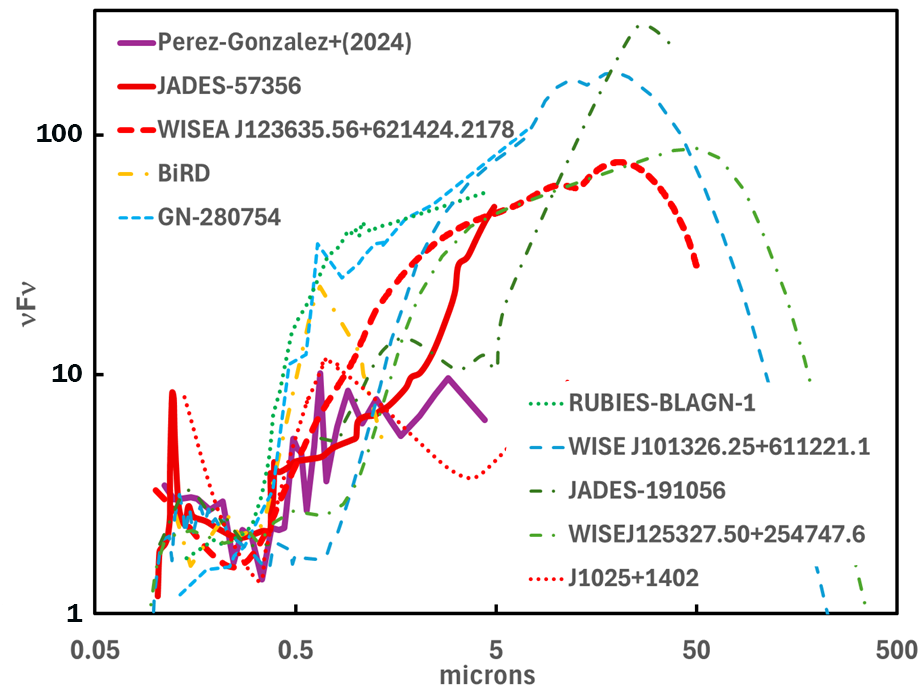}
    \caption{Comparison canonical LRDs with  the nuclear SEDs for several relatively low-redshift LRD analogs { (at rest-frame wavelengths)}. The heavy solid lines show: (1) the average SED from \citet{perez-gonzalez_what_2024}; and (2) the reddest example from the same paper, JADES-57356. WISEA~J123635.56+621424.2  (i.e., {\it the Saguaro} discussed throughout this paper) is the heavy dashed line (the AGN portion only). The others are  BiRD from \citet{loiacono_big_2025}, GN-280754 \citet{juodzbalis_jades_2024}; RUBIES-BLAGN-1 from \citet{wang_rubies_2024}; WISE~J101326.25+611221.1, a particularly luminous bluDOG identified by \citet{toba_discovery_2018}; JADES 191056 \citet{barro_extremely_2024}; WISE J125327.50+254747 \citet{li_searching_2025}; and J1025+1402 from \citet{lin_discovery_2025}.  
}
    \label{fig:lrds}
\end{figure*}


It is not yet clear how to trace LRDs to lower redshifts. For example, a ground-based survey over 3.11 deg$^2$ by \citet{ma_counting_2025} reports a one-order-of-magnitude drop in the LRD number density from $z > 4$ to $1.7 < z < 2.7$, consistent with \citet{kocevski_rise_2024} and \citet{kokorev_census_2024}. In contrast, a wide-area {\it Euclid} survey over 63 deg$^2$ finds numerous LRD candidates at $z < 4$, with number densities increasing down to $z \approx 1.5\text{--}2.5$ before declining at lower redshifts \citep{bisigello_euclid_euclid_2025}. These conflicting results likely reflect fundamental differences in selection methodology. 

For instance, \citet{ma_counting_2025} adopt a stringent cut, $(m_{\rm PSF} - m_{\rm CModel}) < 0.016$, highly sensitive to faint, extended halos. Conversely, the {\it Euclid} team uses a peak–surface-brightness criterion, $\mu_{\max} - m_{\rm pointlike} < -2.6$ mag arcsec$^{-2}$, which remains sensitive to moderate halos but is somewhat less punitive than the criterion from \citet{ma_counting_2025}. The importance of selection criteria is made clear in this section, where we discuss half a dozen proposed low redshift analogs, all of which appear to have the signature V-shaped SED (or one close to this shape) and are compact, but as summarized in Figure~\ref{fig:lrds} exhibit a  wide range of behavior in the longer-wavelength infrared. 

As a baseline, \citet{lin_discovery_2025} identified three LRD analogs at $z\approx0.1$ in the SDSS database, across 9,376 deg$^2$. They note the sample is likely incomplete due to SDSS’s biases against LRDs. Moreover, their selection criteria also differ from those of both \citet{ma_counting_2025} and \citet{bisigello_euclid_euclid_2025}. Nonetheless, they have demonstrated for the first time that low redshift analogs {\it actually exist}! 

The examples found by \citet{lin_discovery_2025} must be rare, given the very large parent sample { (although they recently claimed that their sample is highly incomplete; see discussion in \citealt{lin_lrds2_2026}).} In fact,  canonical LRDs known from existing studies are rare at redshifts less than 4 (e.g., \citealt{kocevski_rise_2024, kokorev_census_2024, ma_counting_2025}). To understand their evolution we need to identify what they become as they age beyond this point. Of course, it is conceivable that they morph into perfectly ordinary-appearing galaxies \citep{khan_where_2025}, in which case following their evolution will be difficult. We take the more optimistic stance that their evolutionary products { may retain some LRD-like properties, albeit modified by subsequent evolution and possibly observed during brief or recurrent accretion episodes.}

We have argued that LRDs could lie in compact, low surface brightness host galaxies at high redshifts. By z $\approx$ 2, compared with z $\approx$ 5, we can expect the hosts to double in diameter and to have much higher surface densities of star formation (e.g., \citealt{genin_dawn_2025}), making the hosts more prominent than for the classical LRDs. We can also expect their interstellar medium (ISM) metallicities to have increased, possibly making their already dust-embedded characteristics even more pronounced. 

With these considerations in mind, we list some sibling candidates and show their SEDs in Figure~\ref{fig:lrds}, along with both an average SED for canonical LRDs and the SED of the reddest LRD known from \citet{perez-gonzalez_what_2024}. All of these proposed “siblings” have a flattening of their SEDs short of 1 $\mu$m, and some have the signature “V-shape”. Whether the rest show this shape  intrinsically depends on the amount of reddening one needs to correct in the observations. Given the increase in dust expected at the lower redshifts, we do not reject any proposed siblings on the basis of a flat, rather than rising, UV SED. 

In addition to {\it the Saguaro} (WISEA J123635.56+621424.2), in Figure \ref{fig:lrds}, we show SEDs of:

\begin{itemize}

\item{The Big Red Dot (BiRD) at $z = 2.33$ 
(\citealt{loiacono_big_2025}). It has a unique SED that turns over toward longer wavelengths near 0.5 $\mu$m.} 

\item{GN 280754 at $z = 2.26$, advertised as {\it the Rosetta stone of JWST-discovered AGN} by \citet{juodzbalis_jades_2024}.}

\item{RUBIES-BLAGN-1 at $z = 3.10$. \citet{wang_rubies_2025} have modeled the SED of this source in detail, finding that the UV is dominated by hot stars that are moderately obscured, with the turnup in the UV reflecting the result of dereddening.}

\item{J101326.25+611220.1 at $z=3.703$ (\citealt{toba_discovery_2018}. In this regard, \citet{noboriguchi_similarity_2023} called attention to the possible relation of DOGs discovered with Spitzer (e.g., \citealt{dey_significant_2008}) and LRDs, particularly for those DOGs with blue excesses (namely, Blue HotDOGs; e.g., \citealt{assef_hot_2016, noboriguchi_extreme_2022}). This galaxy is an excellent example. } 

\item{JADES-GS-191056 at $z = 3.139$ (\citealt{barro_comprehensive_2024}). In the short wavelength JWST/NIRCam bands, this source is extended with a complex morphology, suggesting a possible interaction or other disturbance to its host galaxy as found for a significant number of LRDs by \citet{rinaldi_not_2025}.} 

\item{WISE J125327.50+254747.6, at $z = 0.485$ (\citealt{li_searching_2025}). The compactness of this source is demonstrated in the imaging of \citet{mcgurk_spatially_2015}. This is a Hot DOG, but at an exceptionally low redshift, and with properties suggesting they represent a transient stage of evolution (\citealt{li_searching_2025}).}

\item{{ J1025+1402 is one of three sources identified in SDSS by \citet{lin_discovery_2025} at $z \approx 0.1$ that closely resemble canonical LRDs. This similarity is evident from the close match between its SED and the average SED of canonical LRDs from \citet{perez-gonzalez_what_2024}. These sources therefore provide a unique opportunity to study bona fide low-redshift analogs of LRDs and to investigate their physical nature in much greater detail.}}

\item{\citet{lin_discovery_2024} call attention to some resemblances betweeen LRDs and compact Green Pea AGNs. However the infrared properties of these candidates need to be better constrained before we can classify them with the other LRD siblings. }

\end{itemize}

The following list presents illustrative examples; other LRD siblings likely exist with similar characteristics:

\begin{itemize}
\item {There are a number of galaxies related to J123635.56+621424.2 with similar SEDs and point-like HST images, e.g. IRBG2, IRBG4, IRBG5, and IRBG7 in the notation of  \citet{donley_agn_2010} (their Table 2). These also have extended host galaxies that would fade in surface brightness and probably not be prominent at the redshifts of classical LRDs.}

\item{Some Type II quasars have appropriate SEDs (\citealt{wang_luminous_2025}); better high resolution imaging is needed to confirm that they are sufficiently compact.}

\item{With two entries already in the list above, it is clear that evaluation among BluDOGs and just DOGs is likely to reveal additional candidates (\citealt{noboriguchi_extreme_2022}).}
    
\end{itemize}

The variety and abundance of these “siblings” is promising, indicating that a more thorough search may reveal additional evolutionary counterparts of classical LRDs at later cosmic time.





\section{Discussion and Conclusions}

LRDs have become perhaps the most intensively studied high redshift discovery with JWST.  Two of their distinct characteristics are: (1) point-like morphology { at rest-frame optical wavelengths} \citep[e.g.,][]{labbe_population_2023, baggen_small_2024, matthee_little_2024};  and (2) high number densities, perhaps a few percent of population of galaxies at z $>$ 5 \citep{kokorev_census_2024, kocevski_rise_2024, rinaldi_way_2026}, but dropping steeply from $z = 4$ to $z = 2$ (\citealt{ma_counting_2025}). These behaviors raise a fundamental question: if they lack detectable host galaxies, where did they come from, and, as phrased by \citet{khan_where_2025}, {\it where have all the LRDs gone?} We address the second question first.

There must be some form(s) of galaxy that includes the descendents of the LRDs. Identifying this evolutionary phase requires understanding which types of galaxies resemble LRDs and tracing them back to the redshifts where LRDs were more prevalent. Since it is unclear what direction their evolution will take, it is important to identify a range of possible descendants. Toward that end, we have presented a detailed analysis of the characteristics of {\it the Saguaro}. Through a detailed AGN–host image decomposition and spatial analysis of its NIRSpec spectrum, we reveal a “V-shaped” SED arises from its nuclear region. Although the nucleus lies within a face-on spiral galaxy that is clearly visible at $z \approx 2$, we prove, using a toy experiment, that the surrounding galaxy would be virtually undetectable if observed at $z=7$ due to cosmological surface brightness dimming. This would leave the nucleus entirely “naked”, revealing a very compact, heavily dust-embedded AGN with a characteristic “V-shaped” SED, typical of LRDs. That is, it would readily be identified as a canonical LRD using the limited spectral range provided by the NIRCam filter set at high redshifts. At the moment we are observing it ($z\approx2$), it is a star-forming ULIRG with a strong sub-mm output. Nonetheless, when this fades (in a few tens of Myr), it will resemble LRDs even more closely. Thus, it can take its place among the menagerie of objects at redshifts of $2\text{--}4$ that have been suggested to be related to LRDs. 

The disappearance of the host of {\it the Saguaro} at high redshifts led us to look in more detail for the potential presence of undetected hosts around canonical LRDs. By  stacking the images of the full sample of 99 LRDs from \citet{rinaldi_not_2025}, we found significant excess emission extending to a radius of $0.2$\arcsec, or $\approx2\text{--}3$ kpc in diameter { (when looking at the UV emission)}. That is, on average canonical LRDs at high redshift tend to lie within an enveloping galaxy that is not apparent at typical imaging depths due to surface brightness dimming. { Interestingly, a similar experiment was also performed in the optical by \citet{zhang_unveiling_2025}, who found comparable results, further supporting the presence of faint extended halos in LRDs that likely remain undetected in most NIRCam surveys. An even more compelling case is the recently reported lensed LRD at $z=3.501$ behind Abell S1063 ($\mu\approx2$), identified by \citet{kokorev_deepest_2025} in the GLIMPSE data (\citealt{atek_jwsts_2025}), which reach depths of $\approx30.5\text{--}31$~mag at $5\sigma$. This LRD is hosted by a galaxy with $M_{\bigstar}<10^{7.5}\,M_{\odot}$ and $R_{\rm eff}\approx1$~kpc. Further support comes from the recently identified population of local LRD analogs at $z<1$ (\citealt{park_new_2026, ding_discovery_2026, lin_lrds2_2026}), which also appears, in most cases, to be embedded within extended envelopes, despite the limited spatial resolution of the imaging currently available for these sources.}

Finally, we did an analysis of the general expected effects of surface brightness dimming on LRD hosts as a function of redshift. This shows that most of the host galaxy signal slips below typical detection limits by $z = 7$ and hosts are virtually undetectable by $z = 9$. 
We conclude that current measurements are agnostic as to whether LRDs form in isolation or represent the nuclei of extended host galaxies. This ambiguity likely stems from observational biases: at high redshift, we may be seeing only the bright core, {\it the tip of the iceberg}, while more extended components either remain undetected due to resolution and sensitivity limits or are still in the process of assembling.

In light of the existence of objects (i.e., LRDs) showing a “V-shaped” SEDs from $z \approx 8\text{--}9$ down to $z \approx 0.1$, it becomes clear that the “LRD phase” is not exclusive to the early Universe, but may represent a specific evolutionary stage that galaxies undergo under certain { (and still unknown)} physical conditions. Their compact, red appearance may not indicate a distinct class of objects { (as also recently suggested by \citealt{billand_little_2026})}, but rather a transient and observationally biased view of systems that are more extended and complex than they initially appear, { with the visible component likely tracing the peak of interaction between the host galaxy and the central (still unknown) engine}.

The launch of JWST has undoubtedly transformed our view of the early Universe, greatly enhancing our ability to observe high-$z$ galaxies and fundamentally challenging several aspects of pre-JWST galaxy formation and evolution paradigms. The systematic discovery of LRDs has raised numerous questions about the early growth of black holes and the co-evolution of galaxies and SMBHs at high redshift. While uncovering their true nature is fundamental, it is equally important to investigate what these sources may evolve into at later epochs, as this provides crucial insight into galaxy evolution across cosmic time. { In this context, {\it the Saguaro} may serve as a valuable link, providing a rare test bed in which an LRD-like nucleus can be studied together with its surrounding host galaxy. Future high-resolution spectroscopic follow-up (PID 10311; PIs: P. Rinaldi \& E. Iani) will be essential to spatially isolate the compact core, constrain the physical origin of its V-shaped SED, and assess whether, and to what extent, this system is connected to the broader population of high-redshift LRDs.}.

\begin{acknowledgments}
{ The authors thank the anonymous referee for their constructive comments, which helped improve the manuscript.}
The authors also thank Xiaohui Fan, Xiaojing Lin, Camilla Pacifici, Marianna Annunziatella, Claudia Scarlata, Luis C. Ho, and Pablo~G.\ Pérez\mbox{‑}González for valuable discussions.

\smallskip

This work is based on observations made with the NASA/ESA/CSA JWST. The data were obtained from the Mikulski Archive for Space Telescopes (MAST) at the Space Telescope Science Institute, which is operated by the Association of Universities for Research in Astronomy, Inc., under NASA contract NAS 5-03127 for JWST. These observations are associated with JWST programs GTO \#1181 and GO \#1895.

\smallskip

Some of the data products presented herein were re-
trieved from the Dawn JWST Archive (DJA). DJA is an initiative of the Cosmic Dawn Center (DAWN), which is funded by the Danish National Research Foundation under grant DNRF140.

\smallskip

This research has made use of data obtained from the Chandra Data Archive provided by the Chandra X-ray Center (CXC).

\smallskip

The authors acknowledge the FRESCO team led by PI P. Oesch for developing their observing program with a zero-exclusive-access period. Processing for the JADES NIRCam data release was performed on the lux cluster at the University of California, Santa Cruz, funded by NSF MRI grant AST 1828315. Also based on observations made with the NASA/ESA Hubble Space Telescope obtained from the Space Telescope Science Institute, which is operated by the Association of Universities for Research in Astronomy, Inc., under NASA contract NAS 526555. The data presented in this article were obtained from MAST at the Space Telescope Science Institute. The specific observations analyzed can be accessed via \dataset[DOI: 10.17909/gdyc-7g80, 10.17909/T91019,  10.17909/8tdj-8n28]..

\smallskip

The authors acknowledge use of the lux supercomputer at UC Santa Cruz, funded by NSF MRI grant AST 1828315.

The work of GHR and PR was also supported by  grant
 80NSSC18K0555, from NASA Goddard Space Flight
 Center to the University of Arizona.

AJB acknowledges funding from the "FirstGalaxies" Advanced Grant from the European Research Council (ERC) under the European Union’s Horizon 2020 research and innovation programme (Grant agreement No. 789056)

BER acknowledges support from the NIRCam Science Team contract to the University of Arizona, NAS5-02015, and JWST Program 3215.

The research of CCW is supported by NOIRLab, which is managed by the Association of Universities for Research in Astronomy (AURA) under a cooperative agreement with the National Science Foundation.

SC acknowledges support by European Union’s HE ERC Starting Grant No. 101040227 - WINGS.

YZ, ZJ, BDJ, CNAW, and PL gratefully acknowledge JWST/NIRCam contract to the University of Arizona NAS5-02015.

SA acknowledges support from the JWST Mid-Infrared Instrument (MIRI) Science Team Lead, grant 80NSSC18K0555, from NASA Goddard Space Flight Center to the University of Arizona.

RM, FDE acknowledge support by the Science and Technology Facilities Council (STFC), by the ERC through Advanced Grant 695671 “QUENCH”, and by the UKRI Frontier Research grant RISEandFALL. RM also acknowledges funding from a research professorship from the Royal Society.

LB acknowledges financial support from the Inter-University Institute for Data Intensive Astronomy (IDIA), a partnership of the University of Cape Town, the University of Pretoria and the University of the Western Cape, and from the South African Department of Science and Innovation’s National Research Foundation under the ISARP RADIOMAP+ Joint Research Scheme (DSI-NRF Grant Number 150551) and the CPRR HIPPO Project (DSI-NRF Grant Number SRUG22031677). CJEG acknowledges the financial assistance of the South African Radio Astronomy Observatory
(SARAO) (www.sarao.ac.za).
\end{acknowledgments}





%

\clearpage

\appendix \label{appendix}
\section{AGN-host decomposition}
\begin{figure*}
    \centering
    \includegraphics[width=1.0\linewidth]{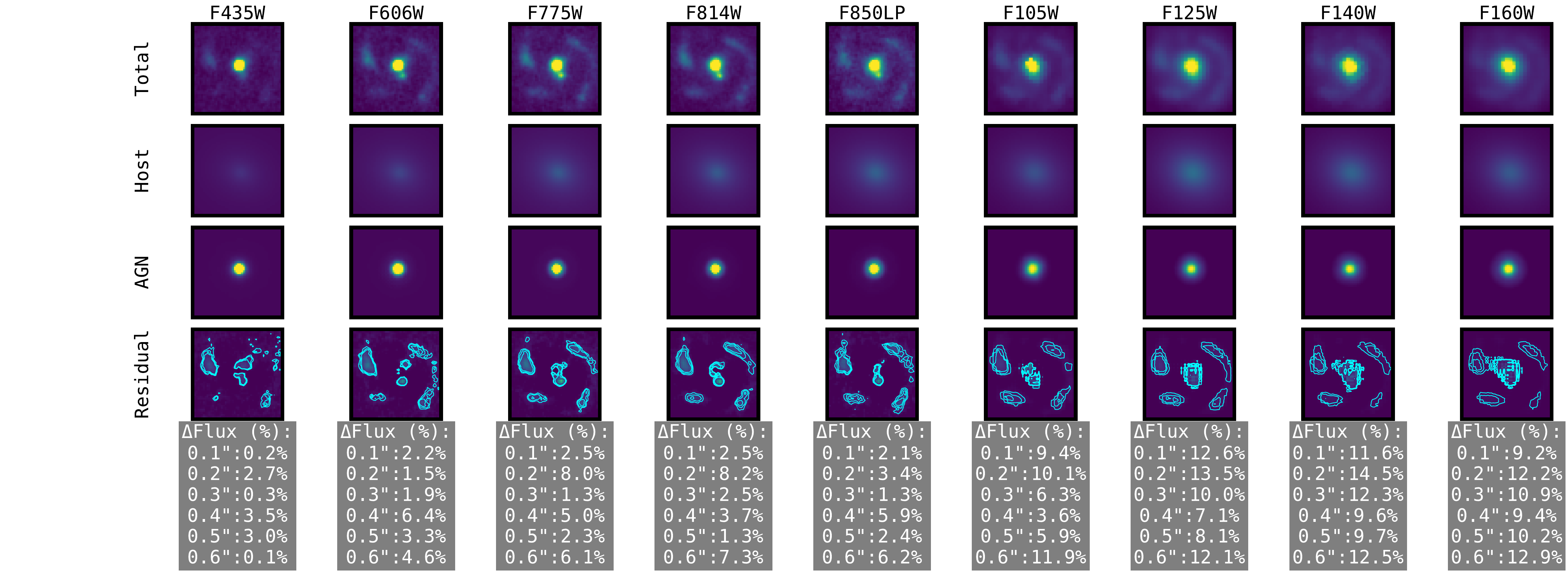}
    \includegraphics[width=1.0\linewidth]{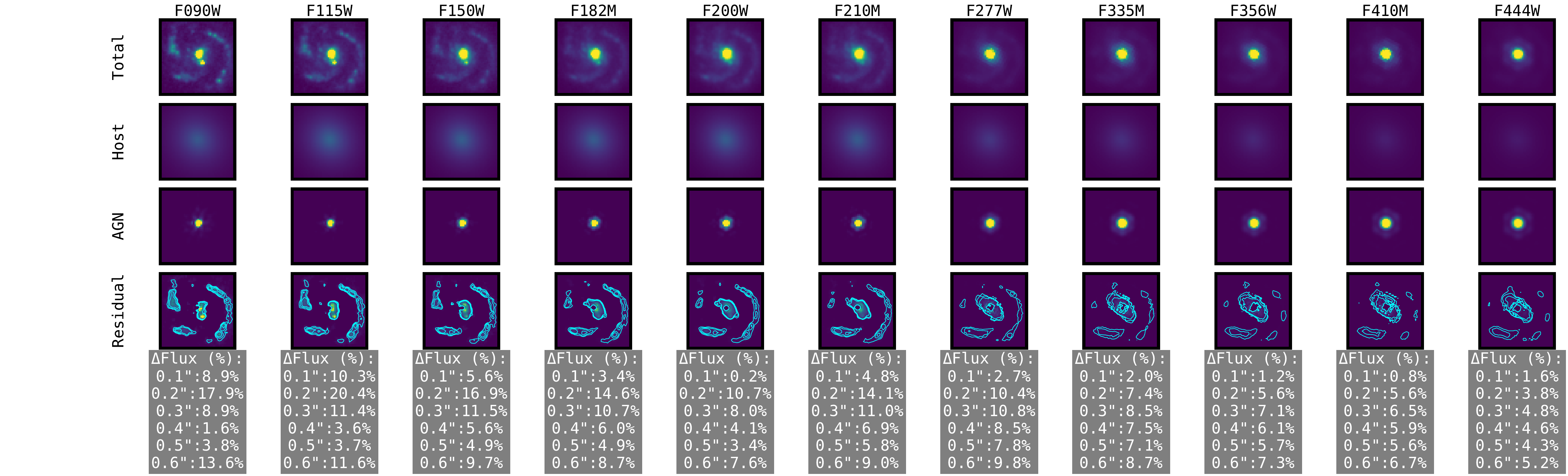}
    \caption{AGN-host image decomposition for HST ({ top}) and NIRCam ({ bottom}). Each cutout is $1.5\arcsec \times 1.5\arcsec$. The first row shows the original image (total), followed by the host and AGN model components in the second and third rows, respectively. The final row displays the residuals, with overlaid contours highlighting the residual structures. All cutouts are shown using the same scale as the original image in each band.
}

    \label{fig:residual}
\end{figure*}

\begin{figure*}
    \centering
    \includegraphics[width=1.0\linewidth]{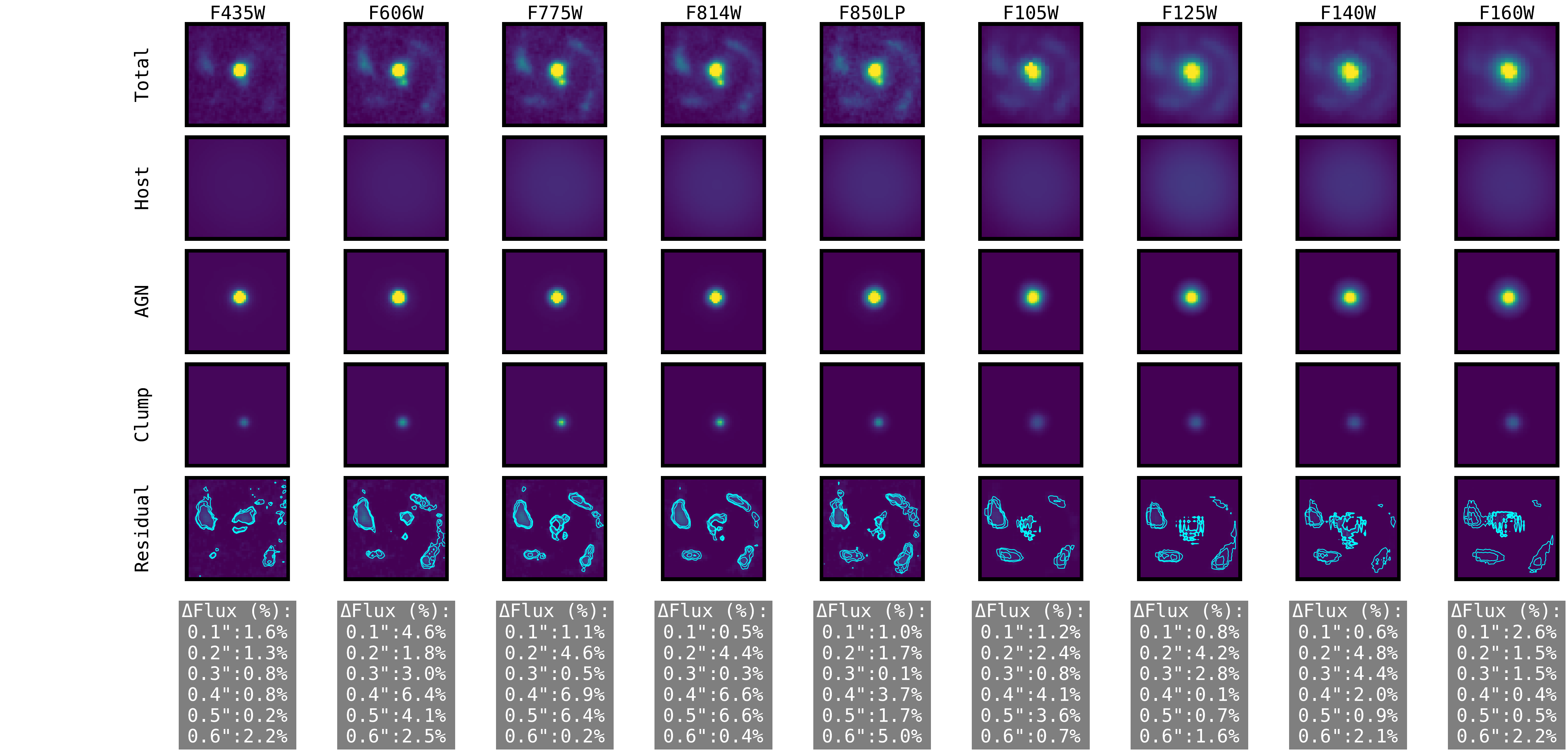}
    \includegraphics[width=1.0\linewidth]{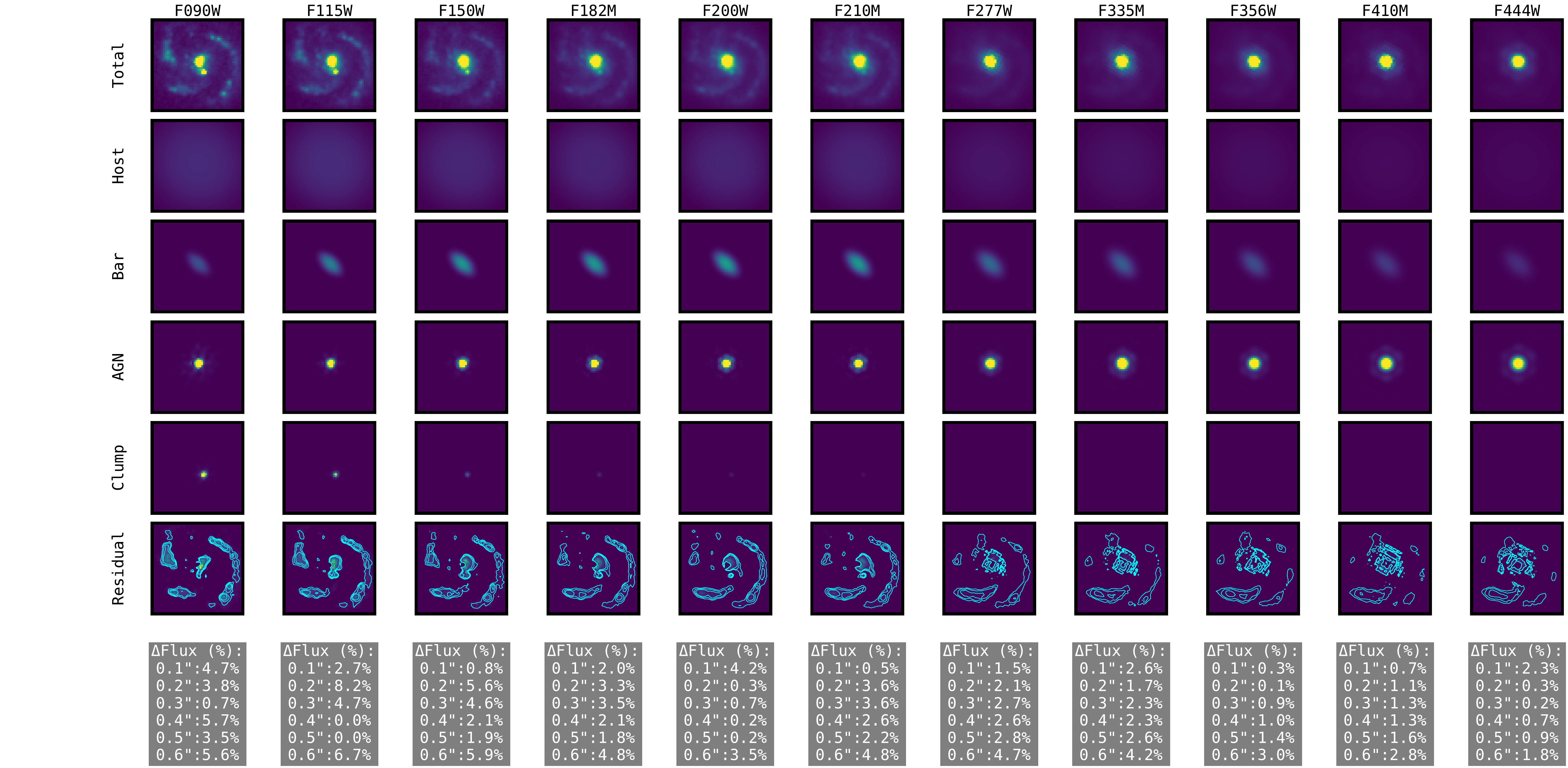}
    \caption{{ AGN--host image decomposition for HST ({ top}) and NIRCam ({ bottom}) using the more complex models that include additional structural components, namely a clump and, where possible, a bar. Each cutout is $1.5\arcsec \times 1.5\arcsec$. For each band, we show the original image (total), the host model, the AGN model, the additional components included in the fit, and the residuals, with contours highlighting residual structures. Owing to the poorer angular resolution of HST, a bar component cannot be robustly fitted in those filters. Similarly, in the NIRCam long-wavelength channel, the broader PSF prevents a reliable fit of the clump component. Although these more complex models provide improved residuals, the added complexity does not lead to significantly different results compared to our baseline decomposition. All cutouts are shown using the same scale as the original image in each band}.
}

    \label{fig:residual_complex}
\end{figure*}

We present the results of the AGN–host image decomposition using both HST and JWST/NIRCam data in Table~\ref{tab:agn_host_fluxes}. We present the data, model, and residual images in $1.5\arcsec \times 1.5\arcsec$ cutouts in Figure~\ref{fig:residual}. We report the residuals in the $\Delta$Flux metric, which quantifies the percentage difference between the model (AGN + host) and the observed flux within the same aperture. This quantity is evaluated across circular apertures with radii ranging from $0.1\arcsec$ to $0.6\arcsec$. Across all HST and JWST bands, the residual flux after AGN-host decomposition remains modest, with average values around 6–7\% and typically below 10\% at all apertures, thus leading to a negligible impact on the photometric measurements for both host and AGN. { In Figure \ref{fig:residual_complex}, we show the results of the more sophisticated {\sc GALFITM} modeling.}

\begin{figure*}
    \centering
    \includegraphics[width=1.0\linewidth]{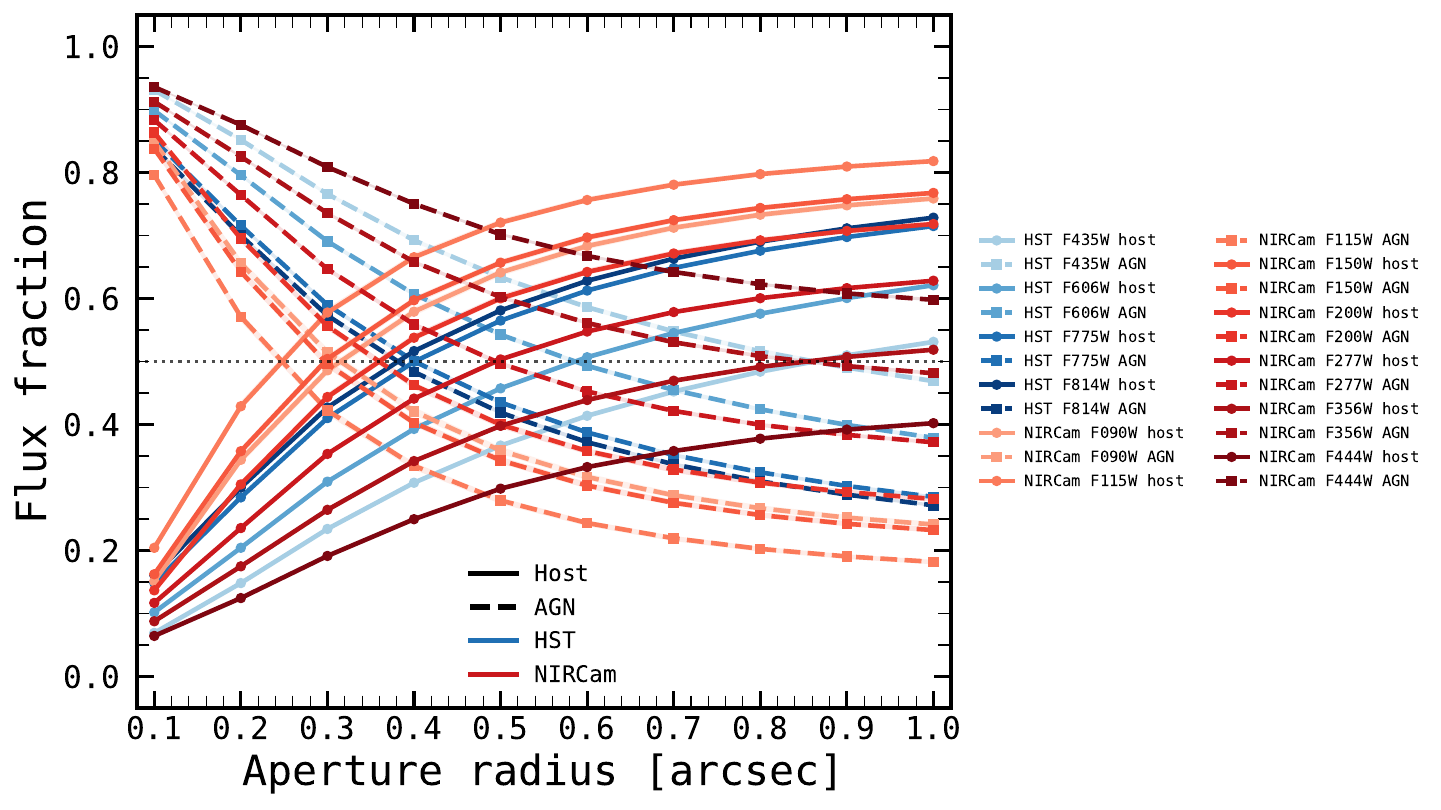}
    \caption{{ AGN and host fractions as a function of radius, shown up to 1\arcsec. The point-like component dominates in the central region, while the host contribution becomes dominant at larger radii.}
}

    \label{fig:flux_fraction_components}
\end{figure*}

{ We also quantify the relative AGN and host contributions as a function of radius. We find that the central regions are dominated by the point-like component, modeled as an AGN, whereas the host becomes dominant at larger radii (Figure \ref{fig:flux_fraction_components}).}

\begin{table*}[ht]
\centering
\caption{Flux densities of the AGN and host galaxy across HST and JWST filters.}
\begin{tabular}{lcc}
\hline
\textbf{Band} & \textbf{AGN Flux ($\mu$Jy)} & \textbf{Host Flux ($\mu$Jy)} \\
\hline
F435W  & 0.42 & 0.54 \\
F606W  & 0.42 & 0.84 \\
F775W  & 0.44 & 1.23 \\
F814W  & 0.47 & 1.45 \\
F090W  & 0.47 & 2.02 \\
F105W  & 0.82 & 2.94 \\
F115W  & 0.82 & 4.80 \\
F125W  & 1.05 & 5.68 \\
F140W  & 1.68 & 7.42 \\
F150W  & 2.17 & 9.03 \\
F160W  & 2.51 & 8.88 \\
F182M  & 3.44 & 13.10 \\
F200W  & 4.97 & 15.30 \\
F210M  & 5.15 & 16.33 \\
F277W  & 12.02 & 23.48 \\
F335M  & 23.33 & 30.48 \\
F356W  & 27.29 & 33.00 \\
F410M  & 44.46 & 37.78 \\
F444W  & 56.49 & 41.57 \\
\hline
\end{tabular}
\label{tab:agn_host_fluxes}
\end{table*}

\section{The stacked UV profile of LRDs with spectroscopic redshifts}
\begin{figure*}
    \centering
    \includegraphics[width=1.0\linewidth]{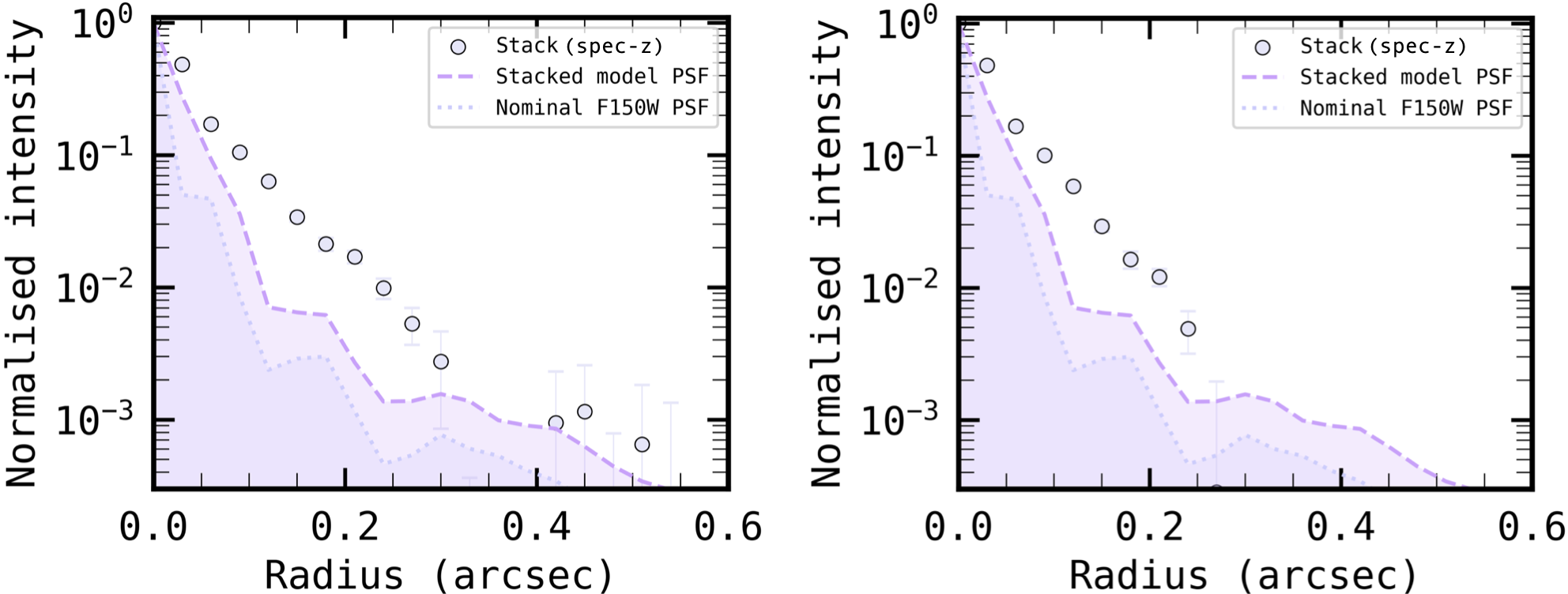}
    \caption{{ Same as Figure~\ref{fig:psf_experiment}, but restricting the stack to LRDs with spectroscopic redshifts and without complex UV morphologies. Despite the smaller sample size, the stacked UV profile remains broader than the simulated stacked PSF, supporting the presence of extended UV emission also in this more conservative subsample.}
}

    \label{fig:psf_experiment_spec}
\end{figure*}

{ In Figure \ref{fig:psf_experiment_spec}, we show the same panel as in Figure~\ref{fig:psf_experiment}, but restricting the analysis to the selected LRDs from \citet{rinaldi_not_2025} with spectroscopic redshifts and without complex UV morphologies. We note that the statistics are limited in this case, comprising less than one third of the full sample. Nevertheless, the extended UV stacked profile is still recovered, further supporting the robustness of this result.}

\clearpage

\facilities{{\sl Chandra}}, {\sl HST}, {\sl JWST}.

\software{\textsc{Astropy} \citep{astropy_collaboration_astropy_2022}, 
\textsc{Bagpipes}
\citep{carnall_vandels_2019},
\textsc{MSAEXP}
\citep{brammer_msaexp_2023}
          \textsc{NumPy} \citep{harris_array_2020},
          \textsc{pandas} \citep{team_pandas-devpandas_2024}
          \textsc{Photutils} \citep{bradley_photutils_2016}, 
          \textsc{TOPCAT} \citep{taylor_topcat_2022},
          \textsc{CIAO} \citep{fruscione_ciao_2006},
          \textsc{XSpec} \citep{arnaud_xspec_1999},
          \textsc{STPSF} \citep{perrin_updated_2014}.
          }

\bibliography{references}{}
\bibliographystyle{aasjournalv7}



\end{document}